\documentclass[11pt, letterpaper]{amsart}

\usepackage{color}
\usepackage[english]{babel}
\usepackage[latin1]{inputenc}
\usepackage{times}
\usepackage[T1]{fontenc}
\usepackage{amsfonts}
\usepackage{epsfig,psfrag,latexsym}
\usepackage{amsmath, amssymb}
\usepackage{graphicx}
\usepackage{paralist}
\usepackage{amsthm}

\newtheorem{theorem}{Theorem}[section]

\newtheorem{lemma}[theorem]{Lemma}
\newtheorem{proposition}[theorem]{Proposition}
\theoremstyle{definition}
\newtheorem{definition}[theorem]{Definition}

\newtheorem{assumption}[theorem]{Assumption}

\theoremstyle{remark}
\newtheorem{remark}[theorem]{Remark}
\newtheorem{example}[theorem]{Example}
\numberwithin{equation}{section}

\newcommand{\reals}{\mathbb R}
\newcommand{\nats}{\mathbb N}

\newcommand{\eps}{\varepsilon}

\newcommand{\such}{\ | \ }
\newcommand{\xpn}[1]{\exp\left(#1\right)}

\newcommand{\prob}{\mathbb{P}}
\newcommand{\qprob}{\mathbb{Q}}
\newcommand{\probalt}[1]{\prob\bra{#1}}
\newcommand{\probnalt}[1]{\prob^{n_k}\bra{#1}}

\newcommand{\esp}{\mathbb{E}}
\newcommand{\espalt}[2]{\esp^{#1}\bra{#2}}

\newcommand{\varalt}[2]{\mathrm{Var}^{#1}\bra{#2}}

\newcommand{\F}{\mathcal{F}}

\newcommand{\M}{\mathcal{M}}
\newcommand{\tM}{\tilde{\mathcal{M}}}
\newcommand{\hM}{\hat{\mathcal{M}}}
\newcommand{\filt}{\mathbb{F}}

\newcommand{\filtrationn}{\filt^n = \pare{\F^n_t}_{0\leq t\leq T}}
\newcommand{\probtriple}{\pare{\Omega, \F, \prob}}
\newcommand{\basis}{\pare{\Omega, \F, \filt, \prob}}
\newcommand{\probtriplen}{\pare{\Omega^n, \F^n, \prob^n}}
\newcommand{\basisn}{\pare{\Omega^n, \F^n, \filt^n, \prob^n}}
\newcommand{\esssup}[2]{\mathop{\textrm{ess sup}}_{#1}\left[ #2\right]}
\newcommand{\essinf}[2]{\mathop{\textrm{ess inf}}_{#1}\left[ #2\right]}

\newcommand{\relent}[2]{H\left(#1\such #2\right)}

\newcommand{\qprnn}{Z^{\qprob,n}}
\newcommand{\qprnno}{Z^{n,0}}
\newcommand{\qprno}{Z^n_1}
\newcommand{\qprnt}{Z^n_2}

\newcommand{\lopital}{l'H\^{o}pital}

\newcommand{\nada}[1]{}

\newcommand{\dfn}{\, := \,}

\newcommand{\Ua}{\mathcal{U}_{\alpha}}
\newcommand{\UUa}{\tilde{\mathcal{U}}_{\alpha}}

\newcommand{\Up}{\mathcal{U}_{p, l}}

\newcommand{\pare}[1]{\left(#1\right)}
\newcommand{\bra}[1]{\left[#1\right]}
\newcommand{\cbra}[1]{\left\{#1\right\}}
\newcommand{\dbra}[1]{[\kern-0.15em[ #1 ]\kern-0.15em]}
\newcommand{\dbraco}[1]{[\kern-0.15em[ #1 [\kern-0.15em[}

\newcommand{\ul}[1]{\underline{#1}}
\newcommand{\ol}[1]{\overline{#1}}

\newcommand{\argmax}[1]{\textrm{argmax}_{#1}}

\title[Large Position Pricing]{Pricing for Large Positions in Contingent Claims}

\author[]{Scott Robertson\\
Carnegie Mellon University}
\address[]{Carnegie Mellon University,
 Department of Mathematical Sciences,
 Wean Hall 6113,
 Pittsburgh, PA 15213, USA}
\email{scottrob@andrew.cmu.edu}
\thanks{The author is supported in part by the National Science Foundation
  under grant number DMS-1312419.}

\date{\today}

\begin{document}




\begin{abstract}
Approximations to utility indifference prices are provided for a contingent
claim in the large position size limit. Results are valid for general
utility functions on the real line and semi-martingale models. It is shown that as the position
size approaches infinity, the utility function's decay rate for large negative
wealths is the primary driver of prices. For utilities with exponential decay,
one may price like an exponential investor. For utilities with a power decay,
one may price like a power investor after a suitable adjustment to the rate at
which the position size becomes large. In a sizable class of diffusion
models, limiting indifference prices are explicitly computed for an
exponential investor. Furthermore, the large claim limit is seen to
endogenously arise as the hedging error for the claim vanishes.

\end{abstract}




\maketitle



\section{Introduction}\label{S:intro}

The last two decades have seen an explosive growth in the financial derivatives market.  Indeed, according to \cite{BIS_Data}, the notional size
of the over-the-counter derivatives market increased from $\$ 94$ trillion in
June of 2000, to $\$ 707$ trillion as of June of 2012.  Similarly,
U.S. issuance of mortgage backed securities (\cite{Mortgage_Data})
increased from $\$ 496$ billion to $\$ 3.2$ trillion between 1996 and 2003
before reverting back to $\$ 1.7$ trillion in 2011.  Due to their complexity,
these contracts are often neither easily traded nor hedged.  The purpose of
this article is to identify a means for pricing such instruments
which takes into account market incompleteness, investor risk aversion, and most
importantly, the large position
size.  In particular, it is sought to identify which aspects of the market and
the investor are the primary drivers of prices.

Let $q$ denote the position size in a derivative contract which the investor holds, but may not trade. The goal is to study the (average bid) utility indifference price $p = p_U(x,q)$ in the
limit that $q\rightarrow \infty$. Here, $U$ is the investor's utility function
and $x$ is the initial
capital. $p$ is defined through the balance equation
\begin{equation}\label{E:uip_def}
u_U(x-qp,q) = u_U(x,0),
\end{equation}
where, given $(x,q)$, the value function $u_U(x,q)$ represents the optimal utility an investor may
achieve by trading in the underlying market. The idea behind indifference pricing traces
back to \cite{HN1989} and the topic has been extensively studied : see
\cite{MR2547456} for a comprehensive review. Clearly, to
compute $p$, knowledge of the value function $u_U$ is crucial.  However,
except for a few special utility functions and models, $u_U$ is not explicitly
known. This presents the primary challenge to obtaining indifference prices
and motivates the study of their approximation.

One approximation occurs in the small claim limit (i.e. as 
$|q|\downarrow 0$).  Here, \cite{MR1491376,MR2233539} obtain first order 
approximations in a Brownian setting, while \cite{MR2011941, MR1802922,
  MR2489605, MR2776692, MR2152255,MR2830428} obtain asymptotic results, regarding both pricing and hedging strategies, for the exponential utility (as well as for general
utilities on the real line in \cite{MR2489605,MR2830428}) in varying degrees
of generality.  In
\cite{MR2288717}, small claim approximations are obtained for utilities
defined on the positive axis.  A key feature present in all these articles is that the market is kept constant as the claim size becomes
small. Also, almost by construction, this approximation
is not appropriate for large investors.

A second approximation occurs by taking a sequence of markets which is
becoming complete in some sense.  In \cite{MR2233539}, asymptotics are provided in a
basis-risk model as the correlation parameter between the hedgeable and
unhedgeable shocks approaches one (this case is treated in detail in Section
\ref{S:basis_risk}).  \cite{MR1398049} obtains results in a Brownian
setting in the case of both fixed and vanishing
portfolio constraints. A key feature of these papers is that as the market
changes, the claim size remains fixed. 

In contrast to the above asymptotics, for large positions is desirable to
allow both the position size and market to vary. This follows by considering
the relationship between owning and hedging a claim. Indeed, as shown in Section \ref{S:basis_risk}, in a Brownian setting with exponential
utility, given the opportunity to purchase claims for an arbitrage free price,
the optimal position size to take (see \cite{MR2212897,
  Siorpaes10}) satisfies the heuristic relationship
\begin{equation}\label{E:opt_rel}
\textrm{risk aversion }\times\textrm{position size} \times \textrm{hedging
  error} \approx \textrm{constant}.
\end{equation}
Thus, for a fixed risk aversion, large position sizes arise in conjunction with vanishing hedging
errors, and hence the market should be allowed to vary.  In fact, through the lens of \eqref{E:opt_rel}, large
claim analysis can be thought of as treating the regime where position size
$\times$ hedging error $\approx$ constant, as opposed to the small
claim or asymptotically complete limits where position size
$\times$ hedging error $\approx 0$. Given
the notional amounts outstanding, the former regime is entirely
consistent with markets where hedging errors are nearly negligible.

Using the above as motivation, for a given sequence of markets $\bold{M}^n$,
claims $h^n$, and respective position sizes $q_n$, the
``large claim limit'' is defined through two
requirements. First (clearly), that $q_n\rightarrow\infty$. Second, that
asymptotically the $\bold{M}^n$ do not permit arbitrage. Note that this
certainly includes the regime when additionally, \eqref{E:opt_rel} holds but
does not require it. For a fixed market, it is shown
in \cite{MR2489605} that no arbitrage is equivalent to $u^n_U(x,0)  <
U(\infty)$, where $u^n_U(x,0)$ is the value function in the $n^{th}$ market
when no claims are held. As $n\uparrow\infty$, the analogous statement is 
$\limsup_{n\uparrow\infty} u^n_U(x,0) < U(\infty)$. This is enforced in
Assumptions \ref{A:val_funct_limit} and \ref{A:val_funct_limit_power} : see Remark
\ref{R:asympt_na} and Proposition \ref{P:val_funct_no_claim}. The idea is that for limiting prices to have any
meaning, the sequence of markets asymptotically should not allow for risk-less profit.

To avoid cumbersome admissibility restrictions, the claims $h^n$ are assumed
to be uniformly bounded and utility functions $U$ on the whole real line are
considered.  The main results of the paper, Theorem \ref{T:uip_big_result} and Proposition
\ref{P:uip_big_result_power} state that the
decay of $U$ for large negative wealths is the primary investor-specific
determinant of the limiting indifference price. For exponential decay, Theorem
\ref{T:uip_big_result} shows prices come together for all utilities $U$ with
the same rate of decay. This theorem has the simple, practical message that an investor with utility
function $U$ should ascertain if there is some $\alpha>0$ such that
$\lim_{x\downarrow -\infty} -(1/x)\log(-U(x)) = \alpha$ (see Examples
\ref{Ex:fund_manager} and \ref{Ex:rep_mkt_maker} for two important, non
exponential utilities), and if so, price like an
exponential investor with risk aversion $\alpha$. Furthermore, under the
additional restriction that $\log(U(x)/(-e^{-\alpha x}))$ remains bounded as
$x\downarrow -\infty$ (see Definition \ref{D:util_funct_class_alt}) Theorem \ref{T:uip_main_result_alt} shows that the total
monetary error incurred by approximating via exponential prices remains bounded, providing a rate of how fast prices come together. 

For utilities with power-like decay the situation is more
complicated. Here, $U$ is assumed to satisfy $\lim_{x\downarrow -\infty}
-U(x)/(-x)^p = 1/l$ for some $p>1$, $l>0$. Proposition \ref{P:uip_big_result_power} shows that prices do come
together for all such $U$, but only after the position size $q_n$ has been
suitably altered to $(-u^n_U(x,0))^{1/p}q_n$. Even though this implies that prices will
typically not come together for the non-adjusted sizes $q_n$, Proposition
\ref{P:uip_big_result_power} still allows for prices to be computed using the
more manageable utility function $U_p(x) = -(1/p)(-x)^p, x\approx -\infty$ (more
precisely, for the convex conjugate function $V_p(y) = ((p-1)/p)y^{p/(p-1)}$:
see equations \eqref{E:indiff_px_formula} and \eqref{E:dual_util_limit_power}). 

In addition to the over-the-counter derivatives market, large
claim limit pricing has applications to the insurance industry.  \cite{BEM_2012} considers large positions pricing for liabilities with both financial and insurancial risks, such as revenue
insurance contracts or mortality derivatives.  Here, the claim is actually the
sum of the contracts.  The crucial assumption 
in \cite{BEM_2012} is that the large claim limit naturally arises, not in
conjuction with vanishing hedging error, but rather with vanishing absolute risk
aversion, as can be deduced by \eqref{E:opt_rel}.

This paper is organized as follows : Section \ref{S:setup} introduces the
general setup for both the models and the family of utility
functions, in the exponential decay case. Section \ref{S:large_claim_limit}
gives the main results for the exponential decay case, as well as providing
examples to highlight the minimality of the given assumptions. Section
\ref{S:power_tails} gives the corresponding results for the power decay case. Using
\cite{MR2094149}, which identifies the value
function for an exponential investor, Section \ref{S:basis_risk} computes limiting prices and monetary
errors in a class of
stochastic volatility, or basis risk, models.  Proposition \ref{P:uip_price_br} shows there are essentially three limiting indifference
prices, depending upon the rate at which $q_n$ becomes large. As will be discussed, this trichotomy of limiting prices appears to be general feature of
large claim analysis, and is intimately related to the theory of Large Deviations. Section
\ref{S:basis_risk} also motivates the relationship in
\eqref{E:opt_rel}, discussing its validity from an equilibrium standpoint using the notion of \emph{partial equilibrium price quantities} introduced
in \cite{MR2667897}. Section \ref{S:basis_risk} concludes by showing that asymptotic market completeness is best defined
through the primal lens of vanishing hedging errors, rather than the dual lens
of collapsing families of martingale measures. Sections \ref{S:more_proofs},
\ref{S:more_proofs_2} and Appendix \ref{S:V_rv_lemmas} contain the proofs.

\section{Setup}\label{S:setup}


\subsection{Assets and Martingale Measures}\label{SS:model}

Let $T>0$ denote the horizon. For each
$n$, let $\basisn$ be a filtered probability space where the filtration $\filtrationn$ satisfies the usual conditions of right-continuity and $\prob^n$
completeness. Assume $\F^n_T = \F^n$ and zero interest rates so that the safe
asset $S_0$ is identically one. The risky
asset $S^n = (S^n_1,\dots,S^n_{d_n})$ is a locally bounded, $\reals^{d_n}$-valued
semi-martingale. In addition to being able to trade in  $S^n$, the investor
owns $q_n$ units of a non-tradable, $\F^n$ measurable, contingent claim $h^n$.

For $n\in\nats$, denote by $\M^{n}$ the set of probability
measures $\qprob^n\ll\prob^n$ on $\F^n$ such that $S^n$ is a local
martingale under $\qprob^n$. Recall that for any $\mu\ll\prob^n$ on $\F^n$ the relative entropy of $\mu$ with respect
to $\prob^n$ is given by $\relent{\mu}{\prob^n} :=
\espalt{\prob^n}{(d\mu/d\prob^n)\log\left(d\mu/d\prob^n)\right)}$. Since it
plays a central role in the analysis below, define:
\begin{definition}\label{D:measure_class} $\tM^{n}\dfn \cbra{\qprob^n\in \M^{n} \ :\ \relent{\qprob^n}{\prob^n} <
  \infty}$.
\end{definition}

Two important examples are:

\begin{example}[Stochastic Volatility with High Correlation]\label{Ex:stoch_vol}

Consider the stochastic volatility model where the asset $S$ and volatility
$Y$ satisfy the stochastic differential equation (SDE)
\begin{equation}\label{E:basis_risk_model}
\begin{split}
\frac{dS^n_t}{S^n_t} &= \mu(Y_t)dt + \sigma(Y_t)(\rho_n dW_t + \sqrt{1-\rho_n^2}dB_t); \qquad dY_t = b(Y_t)dt +
a(Y_t)dW_t;
\end{split}
\end{equation}
where $W,B$ are independent Brownian motions and $\rho_n \in (-1,1)$. When
$h^n = h(Y_T)$ for a function $h$ on the state space of $Y$, these are
alternatively called ``basis-risk'' models. For exponential utility, utility-based
pricing has been extensively studied : see
\cite{MR1926237,MR1491376,MR2094149,MR2048829,MR2431303,MR2648598,MR2547456,MR2395062}
amongst others. Large claim pricing results for this class of models is given
in Section \ref{S:basis_risk}. When $h^n = h(Y_T, S^n_T)$ see \cite{IJS2005,MR2547456} for
utility-based pricing results which are based on Partial Differential Equation
methods.

\end{example}

\begin{example}[Large Markets]\label{Ex:large_market}

Consider, as in \cite{MR2178505}, when the claim is written on a
large market consisting of infinitely many assets, but the investor is
restricted to trading in the only the first $n$ assets.  As an example, let the assets
evolve according to
\begin{equation}
\frac{dS^i_t}{S^i_t} = \alpha^i dt + dW^i_t;\quad i = 1,2,\dots,
\end{equation}
where $W^1,W^2,\dots$ are a sequence of independent Brownian motions and $\sum_{i=1}^{\infty}(\alpha^i)^2 < \infty$. $h$
is a bounded claim measurable with respect to the sigma field
$\sigma(W^1,W^2,...)$.  For concreteness, $h$ can be either an index option
(see \cite{MR3034077}) or a suitably weighted sum of independent claims $h^n$ where $h^n$ is a
function of $S^n$ (see \cite{2006_Milevsky}). Pricing results for this latter
case are briefly discussed in Section \ref{SS:basis_risk_lcl}. 

\end{example}


\subsection{Utility Functions}\label{SS:util_funct}

Throughout the article, a \emph{utility function} will denote any strictly
increasing, strictly concave function $U\in C^2(\reals)$.  The following class
is of particular importance:

\begin{definition}\label{D:util_funct_class} For $\alpha > 0$, denote by $\Ua$ the set of utility functions satisfying $\lim_{x\uparrow\infty} U(x) = 0$ and
\begin{equation}\label{E:u_conv_facts}
\lim_{x\downarrow -\infty} -\frac{1}{x}\log\left(-U(x)\right) = \alpha.
\end{equation}
\end{definition}

The canonical example of $U\in\Ua$ is the exponential utility $U_\alpha$:
\begin{equation}\label{E:exp_alpha}
U_{\alpha}(x) := -\frac{1}{\alpha}e^{-\alpha x}.
\end{equation}
However, $\Ua$ is a richer class of utility functions, as the following
examples show.

\begin{example}[Fund Manager]\label{Ex:fund_manager} Consider the case of
  several investors with respective utilities $U_j \in
\mathcal{U}_{\alpha_j}, \alpha_j > 0$ for $j=1,\dots,J$. As is
typically done in target date retirement funds, assume these investors pool their wealths into a
common fund and delegate a manager to invest the sum.  The manager's
utility function then takes the form $U(x) \dfn \sum_{j=1}^J w_j U_j(x)$,
where $\cbra{w_j}_{j=1}^J$ are the respective weights of each individual
investor.  It readily follows that $U\in\Ua$ for $\alpha \dfn
\max_{j=1,...,J}\alpha_j$.
\end{example}

\begin{example}[Representative Market Maker]\label{Ex:rep_mkt_maker} This
  example concerns the representative market maker from
equilibrium theory, which dates back to \cite{negishi1960welfare} and has been
extensively studied : see \cite{MR1640352, MR1949437, duffie2010dynamic,
  Kramkov_Bank_I} amongst others.  Recall that for a utility function
$U$, the \emph{absolute risk aversion} is defined by
\begin{equation}\label{E:abs_ra_def}
\alpha_U(x) := -\frac{U''(x)}{U'(x)}.
\end{equation}
For $j=1,...,J$, let  $U_j$ be utility functions with
$\lim_{x\uparrow\infty}U_j(x)= 0$, and with risk aversions which
satisfy $i)$ there is a $K_j > 1$ such that $1/K_j \leq \alpha_{U_j}(x) \leq
K_j$ for all $x\in\reals$; and $ii)$
$\lim_{x\downarrow-\infty}\alpha_{U_j}(x) = \alpha_j > 0$. \lopital's
rule implies $U_j\in\mathcal{U}_{\alpha_j}$.  For $v\in
(0,\infty)^J$ the representative market maker's utility is:
\begin{equation}\label{E:rep_mkt_util}
U_v(x)\dfn \sup_{y_1 + ... + y_J = x}\sum_{j=1}^{J}v_j U_j(y_j).
\end{equation}
Here, \cite[Theorem
4.2]{Kramkov_Bank_I} implies for all $v$ that $U_v\in\Ua$ where $\alpha \dfn \left(\sum_{j=1}^J
  (1/\alpha_j)\right)^{-1}$.
\end{example}

For a general $U\in\Ua$, it can be shown that $U$ satisfies the Inada
conditions $\lim_{x\downarrow -\infty}U'(x) = \infty$ and
$\lim_{x\uparrow\infty}U'(x) = 0$ as well as the conditions of Reasonable
Asymptotic Elasticity (see \cite{MR1865021}):
\begin{equation}\label{E:rae}
\liminf_{x\downarrow-\infty}\frac{xU'(x)}{U(x)} >1;\qquad
\limsup_{x\uparrow\infty}\frac{xU'(x)}{U(x)} < 1.
\end{equation}
The normalization $U(\infty) = 0$ is performed only to ensure $\log(-U(x))$ is defined
for all $x\in\reals$. If a utility function $U$ is bounded from above and
satisfies \eqref{E:u_conv_facts} then $U(x)-U(\infty) \in \Ua$. Lastly, for
any utility function $U$ denote by $V$ the convex conjugate to $U$:
\begin{equation}\label{E:util_conv_conj}
V(y) := \sup_{x\in\reals}\cbra{U(x)-xy}.
\end{equation}
It is straightforward to check that for $U\in\Ua$,  $V\in
C^2(0,\infty)$ is strictly convex, can be continuously extended to $0$ by
setting $V(0) = U(\infty) = 0$ and satisfies $\lim_{y\downarrow 0}V'(y) =
-\infty$, $\lim_{y\uparrow\infty} V'(y) = \infty$. Furthermore
\begin{equation}\label{E:dual_util_limit}
\begin{split}
\lim_{y\uparrow\infty} &\frac{V(y)}{V_\alpha(y)} = 1;\qquad V_\alpha(y)\dfn
\sup_{x\in\reals}\cbra{U_\alpha(x)-xy} = 
\frac{1}{\alpha}y(\log(y)-1).
\end{split}
\end{equation}


\subsection{The Value Function and Utility Indifference Price}\label{SS:uip}

Let $n\in\nats$.  A trading strategy $H^n$ is admissible if it is predictable, $S^n$ integrable
under $\prob^n$, and such that the gains process $(H^n\cdot S^n)$ remains above a constant $a$ (which may depend upon $H^n$) almost surely on $[0,T]$. Denote by
$\mathcal{H}^n$ the set of admissible trading strategies. Now, let $\alpha > 0$ and $U\in\Ua$. For $x,q\in\reals$, the value function
$u^n_U(x,q;h^n)$ is defined by
\begin{equation}\label{E:primal_vf}
u^n_U(x,q; h^n) := \sup_{H^n\in\mathcal{H}^{n}}\espalt{\prob^n}{U\left(x
    + (H^n\cdot S^n)_T + q h^n\right)}.
\end{equation}
Note that $n$ appears in two places. The superscript $n$ outside the parentheses
accounts for the dependence of $S^n$ and $\mathcal{H}^n$ upon $n$. The $n$ in $h^n$
  represents the fact that the claim may be changing with $n$. In the case where $h^n\equiv
  0$, set
\begin{equation}\label{E:no_claim_val_funct}
u^n_U(x) := u^n_U(x,q;0).
\end{equation}

Define the average utility indifference (bid) price $p^n_U(x,q;h^n)$ implicitly as the
solution to the equation
\begin{equation}\label{E:uip_price_def}
u^n_U(x) = u^n_U(x - q p^n_U(x,q;h^n), q;h^n).
\end{equation}
Thus, $p^n_U(x,q;h^n)$ is the amount an investor would pay per unit of $h^n$ so
as to be indifferent between owning and not owning $q$ units of the claim.

\begin{remark}\label{R:exp_util_px} For the exponential utility $U_\alpha$ it is well known that
  the indifference price does not depend upon the initial capital (see \cite{MR2152255}).  Thus,
  write $p^n_{U_\alpha}(q;h^n)$ for $p^n_{U_\alpha}(x,q;h^n)$.
\end{remark}


\section{Indifference Prices in the Large Claim
  Limit}\label{S:large_claim_limit}

$p^n_U(x,q_n;h^n)$ is now studied in the limit that
$q_n\rightarrow\infty$. Proofs of all assertions made herein are given in
Section \ref{S:more_proofs}.  The main result states that for
any $x_1,x_2\in\reals$ and $U_1,U_2\in\Ua$, as $q_n\rightarrow\infty$ the
difference between $p^n_{U_1}(x_1,q_n;h^n)$ and $p^n_{U_2}(x_2,q_n;h^n)$ vanishes.  The intuition for this result is gained by inspecting the indifference pricing
formula obtained in \cite[Proposition 7.2 (vi)]{MR2489605}, which is valid in
the current setup:
\begin{equation}\label{E:indiff_px_formula}
p^n_U(x,q_n;h^n) = \inf_{\qprob^n\in\tM^n_V}\left(\espalt{\qprob^n}{h^n} +
  \frac{1}{q_n}\alpha^n_U(\qprob^n)\right),
\end{equation}
where the \emph{entropic penalty functional} $\alpha^n_U$ is given by
\begin{equation}\label{E:penalty_functional}
\alpha^n_U(\qprob^n) \dfn
\inf_{y>0}\frac{1}{y}\left(\espalt{\prob^n}{V\left(y\frac{d\qprob^n}{d\prob^n}\right)}
  + x y
  - u^n_U(x)\right).
\end{equation}
Here, $V$ is from \eqref{E:util_conv_conj}, and $\tM^{n}_V$ is the subset of $\mathcal{M}^n$ such that
$\espalt{\prob^n}{V(d\qprob^n/d\prob^n)} < \infty$.

For $U\in\Ua$, Lemma \ref{L:dual_class_equiv} below implies that
$\tM^n_V = \tM^n$ and hence the variational problem above is
taken over the same set of measures. Furthermore, as $q_n\rightarrow\infty$,
the factor of $(1/q_n)$ in front of $\alpha^n_U(\qprob^n)$ means small values of
$V(z)$ may be disregarded. Therefore, by  \eqref{E:dual_util_limit} one may
replace $V(z)$ with $V_\alpha(z)$ in \eqref{E:penalty_functional}. Calculation
then shows that $\alpha^n_U(\qprob_n) \approx x + (1/\alpha)\left(\log(-\alpha
  u^n_U(x)) + \relent{\qprob^n}{\prob^n}\right)$, and hence
\begin{equation*}
p^n_U(x,q_n;h^n) \approx \frac{1}{q_n\alpha}\log(-\alpha u^n_U(x)) +
\inf_{\qprob^n\in\tM^n}\left(\espalt{\qprob^n}{h^n} +
  \frac{1}{\alpha}\relent{\qprob^n}{\prob^n}\right).
\end{equation*}
Thus, if $\limsup_{n\uparrow\infty}u^n_U(x) < U(\infty) = 0$, the
only part of $p^n_U(x,q_n;h^n)$ dependent upon either $x$ or $U$ vanishes, and 
prices come together.

\subsection{Convergence of Prices}\label{SS:lcl_main_result} The above
argument is made precise under the following assumptions. First, it assumed
that $h^n$ is uniformly bounded in $n$:

\begin{assumption}\label{A:claim} $\|h\|:=\sup_{n}\|h^n\|_{L^{\infty}\probtriplen} < \infty$.
\end{assumption}

The next assumption essentially rules out arbitrage opportunities when
investing in $S^n$, both for each $n$ (see \cite[Assumption
1.4]{MR2489605}) and as $n\uparrow\infty$. Recall, from Section
\ref{SS:model} the definitions of $\tM^n$ and the relative entropy $\relent{\qprob^n}{\prob^n}$.

\begin{assumption}\label{A:val_funct_limit} $\tM^n\neq\emptyset$ for each $n$
  and $\limsup_{n\uparrow\infty}\inf_{\qprob^n\in\tM^{n}}\relent{\qprob^n}{\prob^n}
< \infty$.
\end{assumption}

Regarding Assumption \ref{A:val_funct_limit}, it is well known (see \cite{MR2212897,MR2489605,MR1891730}) that for the exponential utility $U_{\alpha}$,
\begin{equation}\label{E:exp_util_value_function}
u^n_{U_\alpha}(x,q_n;h^n) = -\frac{1}{\alpha}\xpn{-\alpha x -
  \inf_{\qprob^n\in\tM^{n}}\left(\alpha q_n \espalt{\qprob^n}{h^n} + \relent{\qprob^n}{\prob^n}\right)}.
\end{equation}
Now, consider when $q_n\equiv 0$ and Assumption \ref{A:val_funct_limit} does not hold : i.e. for each
$k=1,2,\dots$ there is
an integer $n_k$ such that
$\inf_{\qprob^{n_k}\in\tM^{n_k}}\relent{\qprob^{n_k}}{\prob^{n_k}} >
k$. Since the infimum is strictly bigger than $k$, \eqref{E:exp_util_value_function} implies the existence of an admissible trading strategy $H(n_k)$ such that
\begin{equation*}
\probnalt{(H\left(n_k)\cdot S^{n_k}\right)_T \geq \frac{k}{2\alpha}} \geq
1-e^{-k/2}.
\end{equation*}
Therefore, a very strong form of asymptotic arbitrage holds : namely, there exists a sequence of
admissible trading strategies such that the probability that the terminal wealth
fails to grow like $k$ decreases to $0$ exponentially fast on the
order of $k$.  An asymptotic arbitrage of the form above is similar to a
\emph{strong arbitrage} as defined in \cite{MR2403770}: see this reference for
a more detailed discussion on the topic.

\begin{remark}\label{R:asympt_na} As shown in Proposition \ref{P:val_funct_no_claim} below,
  Assumption \ref{A:val_funct_limit} implies for all $U\in\Ua$ and
  $x\in\reals$ that $\limsup_{n\uparrow\infty} u^n_U(x) < 0 = U(\infty)$.
  This corresponds to the preclusion of asymptotic arbitrage as mentioned in
  the introduction.
\end{remark}

The main result is now presented:

\begin{theorem}\label{T:uip_big_result}

Let Assumptions \ref{A:claim} and \ref{A:val_funct_limit}
hold.  Let $\alpha >0$.  If $q_n\rightarrow\infty$ then for all $U_1,U_2\in\Ua$ and $x_1,x_2 \in
\reals$
\begin{equation}\label{E:price_come_together_alt}
\lim_{n\uparrow\infty}\left|p^n_{U_1}(x_1,q_n;h^n) -
  p^n_{U_2}(x_2,q_n;h^n)\right| = 0.
\end{equation}
\end{theorem}

\begin{remark} Since for any $U\in\Ua$ and $x\in\reals$, $p^n_U(x,q_n;h^n)=
  -p^n_U(x,-q_n; -h^n)$ the convergence in Theorem \ref{T:uip_big_result}
  remains valid for $q_n\rightarrow -\infty$ as well.
\end{remark}


\subsection{Convergence of Total Quantities}\label{SS:total_quant}

Results are now stated which, for $x_1,x_2\in\reals$ and $U_1,U_2\in\Ua$, ensure the total
monetary difference
\begin{equation*}
q_n\left|p^n_{U_1}(x_1,q_n;h^n) - p^n_{U_2}(x_2,q_n;h^n)\right|,
\end{equation*}
remains bounded as $n\uparrow\infty$.  The message is that an investor with
utility function $U\in\Ua$ and initial capital $x\in\reals$ may price as if she were an exponential
investor with risk aversion $\alpha$ and the error in the total amount of money spent by using this approximation remains bounded,
even in the large claim limit.

This type of convergence will not take place
under the general conditions of Theorem \ref{T:uip_big_result} (see Example
\ref{Ex:total_quant_diverge}) and requires
stronger assumptions upon the utility functions. However, under these stronger assumptions it is not necessary for the claim to remain
uniformly bounded as in Assumption \ref{A:claim}. Therefore, assume:

\begin{assumption}\label{A:claim_alt}
For each $n$, $h^n\in L^\infty\probtriplen$.
\end{assumption}

 As for the class of utility functions, convergence results are proved for $\UUa\subset \Ua$ defined by:
\begin{definition}\label{D:util_funct_class_alt}
\begin{equation}\label{E:UUa_def}
\UUa\dfn\cbra{U\in\Ua : 0 < \liminf_{x\downarrow -\infty} \frac{U(x)}{U_{\alpha}(x)} \leq
\limsup_{x\downarrow -\infty}\frac{U(x)}{U_{\alpha}(x)} < \infty}.
\end{equation}
\end{definition}

\begin{remark}

In Example \ref{Ex:fund_manager}, if for 
$j=1,...,J$, $U_j(x) = -(1/\alpha_j)e^{-\alpha_j x}$ then $U\in
\UUa$. Similarly, in Example \ref{Ex:rep_mkt_maker}, if the $U_j$ therein additionally satisfy
$U_j \in \tilde{\mathcal{U}}_{\alpha_j}$ for $j=1,..,J$ then $U\in \UUa$.

\end{remark}

With these definitions and assumptions, total monetary errors are bounded:

\begin{theorem}\label{T:uip_main_result_alt}

Let $\alpha >0$. Let Assumptions \ref{A:val_funct_limit} and
\ref{A:claim_alt} hold. If $q_n\rightarrow\infty$ then for all $U_1,U_2 \in \UUa$ and
$x_1,x_2\in\reals$
\begin{equation}\label{E:price_come_together}
\limsup_{n\uparrow\infty} q_n\left|p^n_{U_1}(x_1,q_n;h^n) -
  p^n_{U_2}(x_2,q_n;h^n)\right| < \infty.
\end{equation}
\end{theorem}

\begin{remark}\label{R:exp_problem}  For the exponential utility price
  $p^n_{U_\alpha}(q_n;h^n)$ of Remark \ref{R:exp_util_px}, consider when Theorem \ref{T:uip_main_result_alt} holds and $p^n_{U_\alpha}(q_n;h^n)$ converges to some limit
  $p_\alpha$.  Even though for $x\in\reals$, $U\in\UUa$ both
  $\lim_{n\uparrow\infty} p^n_U(x,q_n;h^n) = p_\alpha$ and $\limsup_{n\uparrow\infty} q_n|p^n_U(x,q_n;h^n) -
  p^n_{U_\alpha}(q_n;h^n)| < \infty$ hold, it still might be that
  $\lim_{n\uparrow\infty} q_n|p^n_U(x,q_n;h^n) - p_\alpha| = \infty$ (see Proposition
  \ref{P:uip_error_br} for examples).  Here, even though the error in the total
  amount spent by using exponential utility prices (instead of the original
  utility $U$) remains bounded, the error introduced by using the
  \emph{limiting} price for exponential utility tends towards infinity.
\end{remark}

\subsection{Pricing when Only the Position is Changing}\label{SS:market_fixed}

Assume only the position size changes with $n$, i.e.: $\basisn \equiv
\basis$, $S^n \equiv S$ and $h^n \equiv h$. Write $p_U(x,q;h)$ for
$p^n_U(x,q;h)$ and $\tM$ for $\tM^n$. Under Assumptions
\ref{A:claim} and \ref{A:val_funct_limit}, \cite[Proposition 7.5]{MR2489605} proves, for all $U\in\Ua$ and $x\in\reals$:
\begin{equation}\label{E:price_to_min}
\lim_{n\uparrow\infty} p_U(x,q_n;h) =
\inf_{\qprob\in\tM}\espalt{\qprob}{h}.
\end{equation}
For exponential utilities, \eqref{E:price_to_min} has been shown in
\cite{MR2011941,MR2152255}. In fact, for general utilities, \eqref{E:u_conv_facts} is not necessary for this result to hold: 
\cite{MR2489605} proves \eqref{E:price_to_min} follows for all utility functions satisfying the Inada
and Reasonable Asymptotic Elasticity conditions given in Section
\ref{SS:util_funct}. Additionally, by using duality theory for Orlicz spaces
induced by the utility function,  \cite[Proposition 4.2]{MR2830428} proves an
analogous statement to \eqref{E:price_to_min} for non locally bounded
semimartingales under very weak conditions upon the
utility function.  

\subsection{Examples}\label{SS:ex_cex}

Examples are given to highlight the necessity of a) \eqref{E:u_conv_facts} in
Definition \ref{D:util_funct_class} and b) \eqref{E:UUa_def} in Definition
\ref{D:util_funct_class_alt}. Each of the examples considers the one period
trinomial model. Here the filtered space is $\Omega = \cbra{1,2,3}$, $\F = \mathcal{P}^{\Omega}$,
$F_0 = \cbra{\emptyset,\Omega}$, $F_1 = \F$. $S$ and $h$ take the respective values:
\begin{equation}\label{E:trinomial_setup}
\begin{split}
&S_0\equiv 1;\quad S_1(1) = 1+u;\quad S_1(2) = 1;\quad S_1(3) = 1-u,\\
&h(1) = h(3) = h;\quad h(2) = 0.
\end{split}
\end{equation}
where $0 < u < 1$, and $h\neq 0$. Lastly, for each $n$, let $0 < p_n < 1/2$
and define $\prob^n$ by $\prob^n(1)=\prob^n(3) = p_n$, $\prob^n(2)=1-2p_n$. It is clear that Assumptions \ref{A:claim} and
\ref{A:val_funct_limit} hold, and, for any utility function $U$
satisfying the Inada conditions, the indifference price $p^n_U(x,q_n;h)$ satisfies
\begin{equation}\label{E:one_period_trinomial_px}
U(x) = 2p_nU(x-q_np^n_U(x,q_n;h) + q_n h) +
(1-2p_n)U(x-q_np^n_U(x,q_n;h)).
\end{equation}

\subsubsection{On the Necessity of \eqref{E:u_conv_facts}}\label{SSS:necessity}

The first example shows that condition \eqref{E:u_conv_facts} is minimal, at least within the class of
utility functions $U$ such that $\log(\alpha_u))$ is bounded, to guarantee convergence of prices in all
markets satisfying Assumptions \ref{A:claim}, \ref{A:val_funct_limit} where exponential prices converge.

\begin{example}\label{Ex:market_no_conv}
Let $\hat{U}$ be a utility function satisfying i) $\lim_{x\uparrow\infty} \hat{U}(x) = 0$
and ii) for some $K_U>1$, $(1/K_U)\leq \alpha_{\hat{U}}(x)\leq K_U$ for all
$x\in\reals$. Assume that \eqref{E:u_conv_facts} fails : i.e. for some $0 < \ul{\alpha} < \ol{\alpha}$:
\begin{equation}\label{E:u_no_conv_facts}
\ul{\alpha} = \liminf_{x\downarrow\infty} -\frac{1}{x}\log\left(-\hat{U}(x)\right) < \limsup_{x\downarrow -\infty} -\frac{1}{x}\log\left(-\hat{U}(x)\right) = \ol{\alpha}.
\end{equation}
With $\hat{U}^{-1}:(-\infty,0)\mapsto\reals$ as the inverse of $\hat{U}$, there exists $q_n\uparrow\infty$ such that
$-(1/q_n)\hat{U}^{-1}(-e^{q_n})$ does not converge.  In the one-period
trinomial model with $p_n = (1/2)(1-e^{-q_n})$ and $h > (1/\ul{\alpha})$:
\begin{enumerate}[1)]
\item For all $x\in\reals$, $p^n_{\hat{U}}(x,q_n;h)$ does not converge as $n\uparrow\infty$.
\item For all $\alpha > 0$, all $U\in\Ua$ and all $x\in\reals$,
  $\lim_{n\uparrow\infty} p^n_U(x,q_n;h) = \min\cbra{\alpha^{-1},h}$.
\end{enumerate}
\end{example}

\subsubsection{On the necessity of $\UUa$}\label{SSS:total_quant_diverge}

If $U\in\Ua, U\not\in \UUa$ the convergence result in
\eqref{E:price_come_together} may fail for models satisfying Assumptions \ref{A:claim} and \ref{A:val_funct_limit}. The example below considers a utility function where $U(x) = (-1/x)U_\alpha(x)$ for
large negative $x$ : such
a $U$ can easily be constructed and shown to belong to $\Ua$.

\begin{example}\label{Ex:total_quant_diverge}
Let $U\in\Ua$ be such that for some $M>0$, $U(x) =(-1/x)U_\alpha(x)$ if
$x\leq -M$.  Then, in the one period trinomial model with $p_n =
(1/2)(1-e^{-n})$, $h=1$ and $q_n = n^2$, for all $x\in\reals$:
\begin{equation*}
\lim_{n\uparrow\infty} q_n\left(p^n_U(x,q_n;h^n) - p^n_{U_\alpha}(q_n;h^n)\right) = \infty.
\end{equation*}

\end{example}

\subsection{Risk Aversion Asymptotics}\label{SS:connections}

For exponential utility, the absolute risk aversion and the number of units of the claim held are
interchangeable with respect to indifference pricing.  Indeed, for any $q, \alpha > 0$,
$U_\alpha(qx) = qU_{q\alpha}(x)$ for $x\in\reals$ and hence, under Assumptions
\ref{A:claim}, \ref{A:val_funct_limit} (note that $H^n\in\mathcal{H}^{n}\Longleftrightarrow
qH^n\in\mathcal{H}^{n}$):
\begin{equation*}
\begin{split}
p^n_{U_{\alpha}}(q;h^n)&= \frac{1}{\alpha
  q}\log\left(\frac{u^n_{U_{\alpha}}(0)}{u^n_{U_\alpha}(0,q;h^n)}\right) =
\frac{1}{\alpha q}\log\left(\frac{q u^n_{U_{q\alpha}}(0)}{q
    u^n_{U_{q\alpha}}(0,1;h^n)}\right)= p^n_{U_{q\alpha}}(1;h^n).
\end{split}
\end{equation*}

Indifference pricing in the large risk aversion limit has been studied in
\cite{MR1802922, MR1891730} for exponential utilities, and in
\cite{MR2327529,CarRas2011} where results are extended to general utilities on the real line.  Each of these articles show, under suitable hypothesis, that as absolute risk aversion increases to infinity, \emph{if the
  market and claim are not changing as well}, then the (ask) indifference price for
one claim converges the super-replication price of the claim.  This is
entirely consistent with the results in Section \ref{SS:market_fixed} since therein
it is the buyer's indifference price which is considered.


\section{Power Tails}\label{S:power_tails}

Results similar to Theorem \ref{T:uip_big_result} hold for utility
functions with power-like decay for large negative wealths. However, to obtain
convergence of prices, the rate at which $q_n$ becomes large must be suitably
adjusted. This phenomenon is not present for utility functions with
exponential decay.  Proposition \ref{P:uip_big_result_power} below makes the
above statement precise.  To motivate the result, as well as fix notation, 
define the following class of utility functions:

\begin{definition}\label{D:util_funct_class_power}
Let $p>1$ and $l>0$. Define $\Up$ to be the class of utility functions satisfying
$\lim_{x\uparrow\infty} U(x) = 0$ and $\lim_{x\downarrow -\infty} -U(x)/(-x)^p
= 1/l$.
\end{definition}

It can be shown that $U\in\Up$ satisfies both the Inada and Reasonable
Asymptotic Elasticity conditions.  Furthermore, for $U\in\Up$:
\begin{equation}\label{E:dual_util_limit_power}
\lim_{y\uparrow\infty}\frac{V(y)}{V_p(y)} = 1;\qquad V_p(y)\dfn
\hat{l} y^\gamma,
\end{equation}
where $\gamma \dfn p/(p-1)$ is the conjugate exponential to $p$ and
$\hat{l}\dfn (1/\gamma)(l/p)^{\gamma - 1}$. Now, consider the indifference pricing formula from
\eqref{E:indiff_px_formula}. The
factor $(1/q_n)$ again allows one to
disregard small values of $V(z)$ when evaluating $\alpha^n_U(\qprob^n)$ and hence
one may replace $V(z)$ with $V_p(z)$. A lengthy calculation shows for
$U\in\Up$ that $\alpha^n_U(\qprob^n) \approx
(-lu^n_U(x))^{1/p}\ \espalt{\prob^n}{(d\qprob^n/d\prob^n)^\gamma}^{1/\gamma}
$. Substituting the above back in the indifference pricing
formula gives
\begin{equation*}
p^n_U(x,q_n;h^n) \approx \inf_{\qprob^n\in\hat{M}^n}\left(\espalt{\qprob^n}{h^n}
  + \frac{(-lu^n_U(x))^{1/p}}{q_n}\espalt{\prob^n}{\left(\frac{d\qprob^n}{d\prob^n}\right)^\gamma}^{1/\gamma}\right).
\end{equation*}
Here, $\hat{\mathcal{M}}^n$ is the subset of $\mathcal{M}^n$ such that
$\espalt{\prob^n}{V_p(d\qprob^n/d\prob^n)} < \infty$ and the substitution is
allowed because for all $U\in\Up$, $\tM^n_V = \hat{\mathcal{M}}^n$ (see Lemma \ref{L:dual_class_equiv}).  Therefore, by considering prices not for $q_n$, but rather
$q_n(-u^n_U(x))^{1/p}$ one obtains
\begin{equation*}
p^n_U(x,q_n(-u^n_U(x))^{1/p};h^n) \approx \inf_{\qprob^n\in\hat{\M}^n}\left(\espalt{\qprob^n}{h^n}
  + \frac{l^{1/p}}{q_n}\espalt{\prob^n}{\left(\frac{d\qprob^n}{d\prob^n}\right)^\gamma}^{1/\gamma}\right),
\end{equation*}
and, since the right hand side of the above equation does not depend upon $U$
or $x$, prices come together. To formally state the above result, assume, in an analogous
manner to Assumption \ref{A:val_funct_limit}:
\begin{assumption}\label{A:val_funct_limit_power}
With $\hat{\M}^n\dfn \cbra{\qprob^n\in\M^n\such
  \espalt{\prob^n}{(d\qprob^n/d\prob^n)^\gamma} < \infty}$, assume $\hat{\mathcal{M}}^n\neq\emptyset$ for each $n$ and
$\limsup_{n\uparrow\infty}\inf_{\qprob^n\in\hat{\mathcal{M}}^n}
\espalt{\prob^n}{(d\qprob^n/d\prob^n)^\gamma} < \infty$.
\end{assumption}

The main proposition now reads:

\begin{proposition}\label{P:uip_big_result_power}
Let $p>1$ and $l>0$. Let Assumptions \ref{A:claim} and
\ref{A:val_funct_limit_power} hold. If $q_n\rightarrow\infty$ then 
for all $U_1,U_2\in\Up$ and
$x_1,x_2\in\reals$
\begin{equation}\label{E:prices_come_together_power}
\lim_{n\uparrow\infty}\left|p^n_{U_1}(x_1, q_n(-u^n_{U_1}(x_1))^{1/p}; h^n) -
  p^n_{U_2}(x_2, q_n(-u^n_{U_2}(x_2))^{1/p}; h^n)\right| = 0.
\end{equation}
\end{proposition}

\section{Pricing for Stochastic Volatility Models}\label{S:basis_risk}


The results of Section \ref{S:large_claim_limit} are now
specified to the class of models from Example
\ref{Ex:stoch_vol}. Proofs of all assertions made herein are given in
Section \ref{S:more_proofs_2}.


\subsection{Model and Assumptions}\label{SS:basis_risk_m_a}

Let $S^n$ and $Y$ be as in \eqref{E:basis_risk_model}.
Assume that $S^n_0 = 1$ and $h^n = h(Y_T)$ where $h$ is a
function on the state space of $Y$. The probability space $\probtriple$ is two-dimensional Wiener space and the filtration is
the augmented version of the right-continuous enlargement of the natural
filtration $\F^{W,B}$.  Regarding $h$ and the
coefficients $\mu,\sigma,b$ and $a$ in \eqref{E:basis_risk_model}:

\begin{assumption}\label{A:basis_risk}
For $-\infty\leq l < u\leq \infty$, set $E \dfn (l,u)$. $a,b : E\mapsto \reals$ are
continuous and $a^2(y)>0$ for $y\in E$. Furthermore, the SDE for $Y$ in
\eqref{E:basis_risk_model} admits a strong solution with respect to the
$\prob$-augmented filtration of $W$ with $\prob\bra{Y_t\in E, 0\leq t\leq T} =
1$. $\mu,\sigma : E\mapsto\reals$ are measurable such that $\sigma^2(y) > 0,
y\in E$ and 
\begin{equation}\label{E:mkt_px_of_risk}
\lambda(y):=\frac{\mu(y)}{\sigma(y)},
\end{equation}
is bounded on $E$. $h:E\mapsto\reals$ is a continuous and bounded
function. Lastly, $\rho_n \in (-1,1)$ for all $n$.

\end{assumption}

Assumption \ref{A:basis_risk} implies Assumptions
\ref{A:claim}, \ref{A:val_funct_limit} and \ref{A:val_funct_limit_power} for
any $q_n\rightarrow \infty$.  The later two assumptions follow since $\hat{\qprob}^n\in\M^{n}$ for
\begin{equation}\label{E:hat_qn_def}
\frac{d\hat{\qprob}^n}{d\prob} \dfn\mathcal{E}\left(-\int_0^\cdot \rho_n\lambda(Y_t)dW_t - \int_0^\cdot \sqrt{1-\rho_n^2}\lambda(Y_t)dB_t\right)_T,
\end{equation}
and for $\gamma > 1$,
$\sup_{n}\espalt{\prob^n}{(d\hat{\qprob}^n/d\prob)^\gamma} \leq
\textrm{exp}\left(\gamma T \sup_{y\in E}\lambda^2(y)\right)$.
Since it used often below, for $\rho\in\reals$ set
\begin{equation}\label{E:z_rho_def}
Z(\rho)\dfn \mathcal{E}\left(-\rho\int_0^\cdot
  \lambda(Y_t)dW_t\right)_T;\qquad Z\dfn Z(1).
\end{equation}

\begin{example}\label{Ex:br}

The ``basis risk'' case of \cite{MR1926237,MR1491376} treats when
$S^n$ and $Y$ are two geometric Brownian motions with instantaneous
correlation $\rho_n$. This corresponds to $E=(0,\infty)$, $\mu(y) = \mu$, $\sigma(y)= \sigma$,
$b(y) = by$ and $a(y) = ay$ for $\mu,b\in\reals$ and $\sigma, a >
0$.
\end{example}

\subsection{Large Claim Pricing}\label{SS:basis_risk_lcl}

Pricing results for $U\in\Ua$ are now given in the joint limit that
$q_n\rightarrow\infty$ and $\rho_n\rightarrow 1$. As indicated by Proposition \ref{P:br_opt_q}
below, it is convenient to express $q_n$ in terms of $(1-\rho_n^2)^{-1}$ and hence limiting prices are computed
for the following three regimes:
\begin{equation}\label{E:opt_qn}
\qquad q_n = \frac{\gamma_n}{\alpha(1-\rho_n^2)} \quad
\textrm{where}\quad  \begin{cases}\textrm{(i) } \gamma_n\rightarrow 0 \textrm{ but
  }\gamma_n/(1-\rho_n^2)\rightarrow \infty\\
\textrm{(ii) } \gamma_n\rightarrow \gamma > 0\\
\textrm{(iii) } \gamma_n \rightarrow\infty\end{cases}.
\end{equation}

Define $\qprob\sim \prob$ via
$d\qprob/d\prob \dfn Z$ and note that $\qprob$ is the unique martingale
measure in the complete model where $\rho = 1$ and $\F$ is augmented
natural filtration of $W$. 

\begin{proposition}\label{P:uip_price_br} Let $\alpha >0$, $U\in\Ua$ and
  $x\in\reals$.  Let Assumption
  \ref{A:basis_risk} hold. Then,
\begin{equation}\label{E:uip_vals}
\begin{split}
\lim_{n\uparrow\infty} p^n_U(x,q_n;h) =  p_\alpha \dfn \begin{cases} \textrm{(i) : } \espalt{\qprob}{h(Y_T)}\\
\textrm{(ii) : }  -(1/\gamma)\log\espalt{\qprob}{e^{-\gamma h(Y_T)}}\\
\textrm{(iii) : } \essinf{\prob}{h(Y_T)} =
\inf_{y\in E}h(y)\end{cases}.
\end{split}
\end{equation}
\end{proposition}

\begin{remark}\label{R:fast_case} The equality in $(iii)$ follows since $h$ is
  continuous and $\prob\bra{Y_T\in(l',u')} > 0$ for any $(l',u')\in E$. Furthermore, as can be seen in the proof of Proposition
  \ref{P:uip_price_br}, the result for case $(iii)$ holds when $\rho_n$
  is constant and $\gamma_n\rightarrow\infty$. Therefore,
  \eqref{E:price_to_min} implies for each $n$ that
  $\inf_{\qprob^n\in\M^n}\espalt{\qprob^n}{h(Y_T)} = \inf_{y\in E}h(y)$. This
  could also be proved directly using the Martingale Representation Theorem.
\end{remark}

The trichotomy of limiting prices above is motivated by the following
heuristic argument connecting limiting indifference prices to the theory of Large Deviations \cite{MR1619036}.
Assume that $\basis$ and the claim $h$ do not
change with $n$. For each $n$,
assume $h$ decomposes into $h = h_n + Z_n$ where $h_n$ is
perfectly replicable by trading in $S^n$ and $Z^n$ is ``completely
unhedgeable'' in that when pricing $Z_n$, it suffices to assume that one
cannot trade in $S^n$. This implies $u^n_{U_\alpha}(-q_n p_n, q_n;h) =
-(1/\alpha) e^{\alpha q_n p_n} \espalt{}{e^{-\alpha q_n Z_n}}$ and hence
$p^n_{U_\alpha}(q_n;h) =
-1/(\alpha q_n) \log\left(\espalt{}{e^{-\alpha q_n Z_n}}\right)$.

Assume that $h$ is asymptotically hedgeable in such a manner that
$\cbra{Z_n}_{n\in\nats}$ satisfies a Large Deviations Principle (LDP)
\cite{MR1619036} with rate $r_n$ and rate function $I$ where $I(z) = 0
\Leftrightarrow z = 0$.  For the sake of
simplicity, assume further that the $Z_n$ are uniformly bounded, taking values in a
set $E$ and that $I$ is finite on $E$.  Varadhan's
Integral Lemma, \cite[Chapter 4]{MR1619036} in conjunction with
$\lim_{\eps\downarrow 0} \inf_{z\in E}(z + I(z)/\eps) = 0$,
$\lim_{M\uparrow\infty}\inf_{z\in E} (z+ I(z)/M) = \inf_{z\in E} z$ yield
\begin{equation*}
\lim_{n\uparrow\infty} p^n_{U_\alpha}(q_n;h) =  \begin{cases} 0 & q_n/r_n\rightarrow
  0 \\ \inf_{z\in E}\left(z + (1/\alpha)I(z)\right) & q_n/r_n \rightarrow 1\\
\inf_{z\in E}z & q_n/r_n\rightarrow \infty\end{cases}.
\end{equation*}
Thus, if $q_n/r_n\rightarrow 0$ there is no large claim pricing effect since
$Z_n\rightarrow 0$ in Probability and $L^1$. If
$q_n/r_n\rightarrow\infty$ prices converge to the minimum possible value
of the unhedgeable component.  Lastly, if $q_n/r_n\rightarrow 1$ prices are
adjusted via the rate function $I$.  

While the above argument is only motivational, the trichotomy of limiting
prices does hold for other models.  Indeed, in Example \ref{Ex:large_market},
take $T = 1$ and let $h= \sum_{i=1}^\infty \int_0^1 \theta^i_t(dW^i_t + \alpha^i
dt)$ where $\cbra{\theta^i}_{i\in\nats}$ are deterministic functions satisfying
$\sum_{i=1}^\infty \int_0^1 (\theta^i_t)^2dt < \infty$. (Since $h$ is
not bounded, Assumption \ref{A:claim} is violated, however, $h$ does
satisfy the more general assumptions of \cite{MR2489605} and hence the
indifference price is well defined). A direct calculation shows
\begin{equation}\label{E:large_market_price}
p^n_{U_\alpha}(q_n;h) = -
\frac{1}{2}\alpha q_n\sum_{i=n+1}^\infty\int_0^1(\theta^i_t)^2dt  +
\alpha \sum_{i=n+1}^\infty\alpha^i
    \int_0^1\theta^i_tdt.
\end{equation}
Thus, with $r_n \dfn
(\sum_{i=n+1}^\infty \int_0^1(\theta^i_t)^2dt)^{-1}$ the above conclusions
still hold, where the three limiting prices are $i)$ $0$ if
$q_n/r_n\rightarrow 0$, $ii)$ $-(1/2)\alpha l$ if $q_n/r_n
\rightarrow l$ and $iii)$ $-\infty$ if $q_n/r_n\rightarrow
\infty$.

\subsection{Optimal Quantities and Endogenous Large Positions}\label{SS:opt_q_for_p}

The heuristic ``risk aversion $\times$ position size $\times$ hedging error $\approx$ constant'' of
\eqref{E:opt_rel} is now verified in the absence of price
impact. Then, using the notion of \emph{partial-equilibrium price-quantities} (PEPQ)
from \cite{MR2667897}, large positions are shown to endogenously arise even in a
setting where the buyer must find a \emph{single, risk averse} seller (as opposed to the more typical
situation where there is a collection of market makers) in order to purchase the claims.  

Remark \ref{R:fast_case} (the upper
bound follows in a similar manner) implies for each $n$ that the interval of arbitrage free prices for $h(Y_T)$ is
\begin{equation}\label{E:arb_free_h}
I(h) := \left(\inf_{\qprob^n\in \M^n}\espalt{\qprob^n}{h(Y_T)},
  \sup_{\qprob^n\in\M^n}\espalt{\qprob^n}{h(Y_T)}\right) = \left(\inf_{y\in E}h(y), \sup_{y\in E}h(y)\right).
\end{equation}
Let $p_n\in I(h)$ and assume one can buy an arbitrary number of claims for the
price $p_n$. As considered in \cite{2005_DGP, 2007_DGP}, this corresponds to
when buyers can easily find either one another or multiple market makers. A natural problem is to determine
the utility based optimal quantity:
\begin{equation}\label{E:opt_q_for_p}
q_n\in \argmax{q\in\reals} \ u^n_{U}(x-qp^n,q;h).
\end{equation}
For exponential utility, existence of a unique maximizer $q_n$ is proved for
the general framework of Section
\ref{S:large_claim_limit} in  \cite[Theorem 3.1]{MR2212897}. Specified to
the current model, the results in \cite{MR2094149}
enable the precise identification of $q_n$, as well as when it becomes
large. For ease of presentation set $\Lambda\dfn
(1/2)\int_0^T\lambda(Y_t)^2dt$, and, for any $\rho,\gamma\in\reals$ define
\begin{equation}\label{E:g_rho_gamma}
g(\rho,\gamma) \dfn  \frac{\espalt{}{h(Y_T)Z(\rho)e^{-(1-\rho^2)\Lambda-\gamma h(Y_T)}}}{\espalt{}{Z(\rho)e^{-(1-\rho^2)\Lambda-\gamma h(Y_T)}}}.
\end{equation}

\begin{proposition}\label{P:br_opt_q}
Let Assumption \ref{A:basis_risk} hold and let $p_n\in I(h)$. The unique
$q_n$ solving \eqref{E:opt_q_for_p} satisfies $\alpha q_n(1-\rho^2_n)=
\gamma_n$ where $\gamma_n$ is uniquely determined by $p_n =
g(\rho_n,\gamma_n)$. Let $\rho_n\rightarrow 1$. Then, for any subsequence $\cbra{n_k}_{k\in\nats}$
\begin{equation}\label{E:q_n_p_n_lim}
\lim_{k\uparrow\infty} |q_{n_k}| = \infty \Longleftrightarrow
\lim_{k\uparrow\infty}\frac{|p_{n_k}-\hat{p}|}{1-\rho^2_{n_k}} = \infty;\qquad
\hat{p}\dfn \espalt{}{Z(1)h(Y_T)} = \espalt{\qprob}{h(Y_T)}.
\end{equation}
Furthermore, if $p_n\rightarrow p$ for some $p\in I(h)$ then
$\lim_{n\uparrow\infty} \alpha q_n(1-\rho^2_n) = \gamma$ where $\gamma$
uniquely solves $p = g(1,\gamma)$.  $\gamma\neq 0$ if and only if $p\neq
\hat{p}$. 
\end{proposition}

\begin{remark}\label{R:regime}
Proposition \ref{P:br_opt_q} implies that when purchasing optimal quantities,
case $(iii)$ in \eqref{E:opt_qn} never arises.  Case $(i)$ arises if
$p_n\rightarrow \hat{p}$, where $\hat{p}$ is the unique arbitrage free price
in the complete model.  For all other limiting prices $p$, case $(ii)$
arises. 
\end{remark}

Proposition \ref{P:br_opt_q} implies the heuristic in \eqref{E:opt_rel},
provided for each $n$ one may buy claims at a price $p\neq\hat{p}$. At first glance, it may seem unrealistic that one
could engage a seller at this price: however this is
indeed possible in the setting of PEPQ from \cite{MR2667897}. To define a
PEPQ, some additional notation is needed. Let
$X,X'$ be two bounded, $\F_T$ measurable random variables.  For the
exponential utility $U_\alpha$, denote by
$u^n_{\alpha}(q_n;h,X)\dfn u^n_{U_\alpha}(0,q_n; h + X/q_n)$ the value
function for holding $q_n$ claims of $h(Y_T)$ and one unit of $X$.  Now, consider a
second exponential investor with risk aversion $\delta > 0$. A pair
$(q_n,p_n)$ where $p_n\in I(h)$ is called a PEPQ in the $n^{th}$ market if 
\begin{equation*}
q_n\in\argmax{q\in\reals}\left(e^{\alpha q_n p_n}u^n_\alpha(q_n;h,X)\right);\qquad
  q_n \in \argmax{q\in\reals}\left(e^{-\delta q_n p_n} u^n_\delta(-q_n; h, X')\right).
\end{equation*}
In other words, $(q_n,p_n)$ is a PEPQ if, for the price $p_n$, it is optimal for the $\delta$ risk averse investor to sell
$q_n$ units of $h(Y_T)$ and for the $\alpha$ risk averse investor to buy $q_n$
units of $h(Y_T)$.  As shown in \cite[Theorem 5.8, Remark 5.0, Corollary 3.16]{MR2667897}, if
$\alpha X - \delta X'$ is not replicable then there exists a unique PEPQ with $q_n\neq 0$, otherwise there is no PEPQ. This is the reason why the
additional endowments $X,X'$ must be added.

Now, let $X \equiv 0$ and assume the seller holds a position in $h$ consistent
with \eqref{E:opt_rel}: i.e. $X' = (\delta \gamma / (1-\rho^2_n)) h(Y_T)$ for some
$\gamma > 0$.  Since $h(Y_T)$ is not replicable, it follows that a PEPQ
$(p_n,q_n)$ exists.  Furthermore, $(p_n,q_n)$ must satisfy the optimality
conditions  (recall \eqref{E:g_rho_gamma} for $q_n = \gamma_n / (1-\rho^2_n)$):
\begin{equation*}
p_n = g(\rho_n, \alpha \gamma_n) = g(\rho_n,\delta(\gamma-\gamma_n)).
\end{equation*}
It clearly holds that $\gamma_n =\gamma\delta/(\delta+\alpha)$ and $p_n$
satisfies $p_n =
g(\rho_n, \gamma\alpha\delta/(\alpha+\delta))$. Such a $p_n$ exists by Lemma \ref{L:h_funct_facts} below, and, as
$\rho_n\rightarrow 1$ it follows that $p_n\rightarrow p\neq \hat{p}$ where $p
= g(1, \gamma\alpha\delta/(\alpha+\delta))$. Thus, with both buyer and seller
acting optimally, the buyer enters into the regime of \eqref{E:opt_rel} and
the seller is willing to sell for a price $p_n\approx p \neq
\hat{p}$. The message is that as long as there exists a single investor in the regime of \eqref{E:opt_rel}, whether or
not she has entered it optimally, it possible for other investors,
acting optimally, to
enter into the regime \eqref{E:opt_rel} as well. Given
the actual notional sizes existing in the market, it is entirely reasonable to
assume some investor is in the regime of \eqref{E:opt_rel}.

\subsection{Monetary Errors} It is of interest to
know when the monetary error, introduced by using the limiting exponential
utility price, remains bounded. The following Proposition identifies precise
conditions on the $\gamma_n$ from \eqref{E:opt_qn} when this is the
case. Then, assuming for each $n$ that one can buy claims for a fixed $p\in
I(h)$ it is shown that if one buys the optimal number of claims as in
Proposition \ref{P:br_opt_q} so that $\gamma_n$ satisfies $p=
g(\rho_n,\gamma_n)$, then monetary errors are always bounded. For
the sake of brevity, case $(iii)$ is excluded.

\begin{proposition}\label{P:uip_error_br} Let $\alpha > 0$. Let Assumption
  \ref{A:basis_risk} hold.   For $q_n$ from \eqref{E:opt_qn} and
  $p_\alpha$ from Proposition
  \ref{P:uip_price_br}, as $\rho_n\rightarrow 1$:
\begin{equation}\label{E:uip_error_val}
\begin{split}
\limsup_{n\uparrow\infty} &\ q_n\left|p^n_{U_\alpha}(q_n ;
  h)-p_\alpha\right|<\infty \Longleftrightarrow \begin{cases} \textrm{(i)
    : }  \limsup_{n\uparrow\infty}\frac{\gamma_n^2}{1-\rho_n^2} < \infty \\ \textrm{(ii) : }
  \limsup_{n\uparrow\infty}\frac{|\gamma_n-\gamma|}{1-\rho_n^2} < \infty \end{cases}.
\end{split}
\end{equation}
Furthermore, if $\gamma_n$ is chosen optimally as in Proposition \ref{P:br_opt_q} for
a fixed $p\in I(h)$ : i.e. $\gamma_n$ satisfies $p=g(\rho_n,\gamma_n)$ then
monetary errors are always bounded.
\end{proposition}

\subsection{On the Optimal Hedging Strategy}

Assume $q_n$ takes the form in \eqref{E:opt_qn}. As shown in the proof of Proposition \ref{P:uip_mu_facts} below, for an
exponential investor, the optimal hedging strategy $\hat{\pi}^n$ \footnote{If
  the class of admissible trading strategies is enlarged to include strategies
  such that the resultant wealth process is a $\qprob^n$ supermartingale for
  all $\qprob^n\in\tM^n$ : see \cite{MR1891731, MR2489605}} satisfies $\hat{\pi}^n_t = (1/(\alpha\sigma(Y_t)))\left(\lambda(Y_t) +
  (\rho_n/(1-\rho^2_n))\theta^n_t\right)$ where $\theta^n$ satisfies (recall
$\Lambda = (1/2)\int_0^T\lambda(Y_t)^2dt$):
\begin{equation}\label{E:mart_rep_opt_strat}
\mathcal{E}\left(\int_0^\cdot
  \theta^n_t(dW_t + \rho_n\lambda(Y_t)dt)\right)_T =
\frac{e^{-(1-\rho^2_n)\Lambda - \gamma_n
    h(Y_T)}}{\espalt{}{Z(\rho_n)e^{-(1-\rho^2_n)\Lambda-\gamma_n h(Y_T)}}}.
\end{equation}
That such a $\theta^n$ exists follows from the Martingale Representation
Theorem.  Now, if $\gamma_n\rightarrow \gamma \neq 0$ then under the given
hypothesis on the model coefficients, $\theta^n\rightarrow \theta$ in the
sense that $\lim_{n\uparrow\infty} \espalt{}{\int_0^T(\theta^n_t -
  \theta_t)^2dt} = 0$, where
$\theta$ solves \eqref{E:mart_rep_opt_strat} at $\rho_n = 1$.  Thus, even though  $\hat{\pi}^n$ is taking ever larger (in magnitude) positions in the risky
asset $S^n$, the normalized trading strategy $\hat{\pi}^n/q_n$ converges to $\hat{\pi} \dfn
\theta/(\gamma\sigma)$. Then, using \eqref{E:mart_rep_opt_strat} again it
follows that for $p_\alpha$ as in Proposition \ref{P:uip_price_br}, $(p_\alpha,\pi)$ is a super-hedge in the complete model in that at $\rho=1$
\begin{equation*}
-p_\alpha + h(Y_T) +\int_0^T\hat{\pi}_t\frac{dS_t}{S_t} =
\frac{1}{2\gamma}\int_0^T\theta^2_tdt.
\end{equation*}
Furthermore, by
the very definition of the indifference price, for the
non-normalized strategy $\hat{\pi}^n$:
\begin{equation*}
\espalt{}{e^{-\alpha(\int_0^T \hat{\pi}^n_t
    dS^n_t/S^n_t + q_n(h(Y_T)-p^n_{U_\alpha}(q_n;h))}} = -\alpha u^n_{U_\alpha}(0) =
\espalt{}{Z(\rho_n)e^{-(1-\rho^2_n)\Lambda}}^{1/(1-\rho^2_n)}\leq 1,
\end{equation*}
where the second equality above comes from \eqref{E:basis_risk_model_vf} below.
Thus, $\hat{\pi}^n$ and the initial capital $-q_np^n_{U_\alpha}(q_n;h)$ provide a robust ``super hedging'' strategy in that for any
constant $C >0$
\begin{equation*}
\sup_{n}\prob\bra{\int_0^T \hat{\pi}^n_t
    dS^n_t/S^n_t + q_n (h(Y_T)-p^n_{U_\alpha}(q_n;h)) \leq -C} \leq e^{-\alpha C}.
\end{equation*}

\subsection{Asymptotic Completeness and the Local Martingale Measures}

Though $\rho_n\rightarrow 1$, the family of local martingale measures
$\tM^n$, even when restricted to $\F^W$, is
not collapsing to a singleton with respect to the weak convergence of
probability measures.  This follows immediately from \eqref{E:arb_free_h}
since $h(Y_T)$ is $\F^W$ measurable.  Indeed, setting $\qprob^n_W\dfn
\qprob^n\big|_{\F^W}$, \eqref{E:arb_free_h} implies for all $n$ that
\begin{equation*}
\sup_{\qprob^n\in\tM^{n}}\espalt{\qprob^n_W}{h(Y_T)} = \sup_{y\in
  E}h(y);\qquad \inf_{\qprob^n\in\tM^{n}}\espalt{\qprob^n_W}{h(Y_T)} =
\inf_{y\in E} h(y).
\end{equation*}
Therefore, it cannot be that for \emph{any two} sequences of measures $\qprob^{n,1},
\qprob^{2,n}\in\M^n$ that
$\lim_{n\uparrow\infty}|\espalt{\qprob^{n,1}_W}{h(Y_T)} -
\espalt{\qprob^{n,2}_W}{h(Y_T)}| = 0$. The next proposition, which finishes
the section, reinforces this
fact, as well as provides an alternate description of the difference between
the limiting indifference and traded prices in terms of the
relative entropy of two sequences of local martingale measures in $\tM^n$. 

\begin{proposition}\label{P:uip_mu_facts} Let Assumption \ref{A:basis_risk} hold. Let
  $\rho_n\rightarrow 1$, $p\in I(h)$, $p\neq\hat{p}$. Let $q_n,\gamma$ be as
  in Proposition \ref{P:br_opt_q} and $p_\alpha$ be as in case $(ii)$ of Proposition \ref{P:uip_price_br}. Then
\begin{equation}\label{E:two_rel_ent_comp}
\lim_{n\uparrow\infty}\relent{\qprob^{n,q_n}_W}{\hat{\qprob}^n_W} =
\gamma(p_\alpha - p),
\end{equation}
where $\qprob^{n,q_n}\in \tM^{n}$ solves the dual problem
  \eqref{E:exp_util_value_function} with $q_n$ and $\hat{\qprob}^n\in\tM^{n}$ is from \eqref{E:hat_qn_def}.
\end{proposition}


\section{Proofs from Sections \ref{S:large_claim_limit} and \ref{S:power_tails}}\label{S:more_proofs}

\subsection{Preliminaries} Unless otherwise stated, all expectations within this
  section are taken with respect to $\prob^n$ and denoted by $\esp^n$. For any $\qprob^n\ll
  \prob^n$ write $\qprnn \dfn (d\qprob^n/d\prob^n)|_{\F^n}$.

Let $\alpha > 0$, $U\in\Ua$ and define $V$ as in
\eqref{E:util_conv_conj}. As in Section \ref{S:large_claim_limit},
define $\tM^{n}_V$ as the set of $\qprob^n\in\M^{n}$ such that
$\espalt{n}{V(d\qprob^n/d\prob^n)} < \infty$.  By applying Lemma \ref{L:dual_class_equiv} with
$Y = Z^{\qprob,n}$ and $y=1$ it follows that $\tM^{n}_V = \tM^{n}$ for all
$U\in\Ua$.   Therefore, the indifference pricing formula
\eqref{E:indiff_px_formula} specifies to
\begin{equation}\label{E:util_indif_price_rep_pos}
\begin{split}
p^n_U&(x,q;h^n)=\inf_{\qprob^n\in\tM^{n}}\left(\espalt{n}{
      h^nZ^{\qprob,n}} + \frac{1}{q}\alpha^n_U\left(\qprob^n\right)\right).
\end{split}
\end{equation}
In a similar manner, Lemma \ref{L:dual_class_equiv} implies $\tM^n_V =
\hat{\mathcal{M}}^n$ for all $U\in\Up$ and hence
\begin{equation}\label{E:util_indif_price_rep_pos_power}
\begin{split}
p^n_U&(x,q;h^n)=\inf_{\qprob^n\in\hat{\mathcal{M}}^n}\left(\espalt{n}{
      h^nZ^{\qprob,n}} + \frac{1}{q}\alpha^n_U\left(\qprob^n\right)\right).
\end{split}
\end{equation}


\subsection{Proofs}

The proofs of Theorems \ref{T:uip_big_result} and Proposition
\ref{P:uip_big_result_power} follow two steps:
\begin{enumerate}[1)]
\item Verify that Assumptions \ref{A:val_funct_limit} and \ref{A:val_funct_limit_power} imply the "no asymptotic arbitrage" condition  $\limsup_{n\uparrow\infty} u^n_U(x) < U(\infty) = 0$ for all $U\in\Ua$ and $U\in\Up$ respectively.
\item For $V_\alpha$, $V_p$ as in \eqref{E:dual_util_limit} and
  \eqref{E:dual_util_limit_power} respectively, rigorously justify the substitutions of $V$ with $V_\alpha$ (for $U\in\Ua$) and $V$ with $V_p$ (for $U\in\Up$) when computing the penalty functionals $\alpha^n_U(\qprob^n)$.
\end{enumerate}

\begin{proposition}\label{P:val_funct_no_claim} Let $\alpha>0$, $p>1,l>0$ and
  $x\in\reals$. Then Assumption \ref{A:val_funct_limit} implies
  $\limsup_{n\uparrow\infty} u^n(x) < 0$ for $U\in\Ua$.  Similarly, Assumption
  \ref{A:val_funct_limit_power} implies $\limsup_{n\uparrow\infty} u^n_U(x) <
  0$ for $U\in\Up$.
\end{proposition}

\begin{proof}[Proof of Proposition \ref{P:val_funct_no_claim}]

In view of Assumptions \ref{A:val_funct_limit} and
\ref{A:val_funct_limit_power} there exist sequences of measures
$\qprob^n_1,\qprob^n_2 \in
\M^n$ and a constant $C>0$ so that
\begin{equation}\label{E:relent_ub}
\sup_{n}\espalt{n}{V_\alpha(d\qprob^n_1/d\prob^n)} \leq C \ \textrm{ (Ass.
  \ref{A:val_funct_limit})};\qquad \sup_n\espalt{n}{V_p(d\qprob^n_2/d\prob^n)}
\leq C\ \textrm{ (Ass. \ref{A:val_funct_limit_power})}.
\end{equation}
Write $\qprno \dfn d\qprob^n_1/d\prob^n$ and $\qprnt\dfn d\qprob^n_2/d\prob^n$.
For $x\in\reals$ it follows from \cite{MR2489605} that
\begin{equation}\label{E:no_claim_ub}
u^n_U(x) \leq \inf_{y > 0}\left(\espalt{n}{V\left(y\qprno\right)} +
  xy\right) \ (U\in\Ua);\qquad u^n_U(x) \leq \inf_{y > 0}\left(\espalt{n}{V\left(y\qprnt\right)} +
  xy\right) \ (U\in\Up).
\end{equation}
The argument below is nearly identical for $U\in\Ua$ and
$U\in\Up$: thus, it will be given for $U\in\Ua$ and only the adjustments needed for
$U\in\Up$ will be mentioned. Applying Lemma \ref{L:some_Y_V_facts} with $Y=\qprno$ shows there is a unique
$y_n > 0$ solving the minimization problem in \eqref{E:no_claim_ub} and the first order conditions are $x =
-\espalt{n}{\qprnno V'(y_n\qprno)}$.  Assume, for now, that
\begin{equation}\label{E:y_n_liminf}
\liminf_{n\uparrow\infty} y_n > 0.
\end{equation}
Using the first order conditions for $y_n$:
\begin{equation*}
u^n_U(x) \leq \espalt{n}{V\left(y_n\qprno\right)} +
  xy_n = -\espalt{n}{\left(y_n\qprno V'(y_n\qprno) - V(y_n\qprno)\right)}.
\end{equation*}
Set $f(z) \dfn zV'(z) - V(z)$.  Note that $f'(z) = zV''(z) > 0$ and 
$\lim_{z\downarrow 0}f(z) = 0$, since $U(\infty) = 0$. In view of \eqref{E:y_n_liminf}, take $\delta > 0$
such that $y_n \geq \delta$ for large $n$.  Since $f$ is increasing and non-negative
\begin{equation*}
u^n_U(x) \leq - \espalt{n}{f(y_n\qprno)} \leq -\espalt{n}{f(\delta\qprno)}\leq 0.
\end{equation*}
Assume, by way of contradiction, there exists a sequence (still labeled $n$) such that
$\lim_{n\uparrow\infty} u^n_U(x) = 0$. The above inequality implies
$\lim_{n\uparrow\infty}\espalt{n}{f(\delta\qprno)} = 0$, and hence for all $\eps > 0$ that
$\lim_{n\uparrow\infty}\prob^n\bra{\qprno\geq \eps} = 0$. In view of
\eqref{E:relent_ub} the $\qprno$ are ``uniformly integrable'' in that $\lim_{\lambda\uparrow\infty}\sup_n\espalt{n}{\qprno 1_{\qprno\geq
    \lambda}} = 0$.  This follows  because for all $z>0$ and
$\lambda > 1$
\begin{equation*}
z1_{z\geq\lambda} \leq \frac{\lambda}{V_\alpha(\lambda)}\left(V_\alpha(z) +
  \frac{1}{\alpha}\right);\qquad \left(\textrm{resp. } x1_{x\geq\lambda} \leq \frac{\lambda}{V_p(\lambda)}V_p(z)\right),
\end{equation*}
and because $\lim_{\lambda\uparrow\infty} V_\alpha(\lambda)/\lambda = 0$
(resp. $\lim_{\lambda\uparrow\infty} V_p(\lambda)/\lambda = 0$). Now, fix $\eps > 0$ and choose $\lambda$ so large that
$\sup_n\espalt{n}{\qprno 1_{\qprno\geq \lambda}} \leq \eps$. Since
$\qprno\in\tM^n$,
\begin{equation*}
\begin{split}
1 &= \espalt{n}{\qprno} = \espalt{n}{\qprno\left(1_{\qprno\leq \eps} +
    1_{\eps < \qprno < \lambda} + 1_{\qprno\geq \lambda}\right)} \leq \eps  + \lambda\prob^n\bra{\qprno > \eps} + \eps.
\end{split}
\end{equation*}
Taking $n\uparrow\infty$ and then $\eps\downarrow 0$ gives a contradiction and
hence the result holds assuming \eqref{E:y_n_liminf}.

To prove \eqref{E:y_n_liminf}, recall that $y_n$ satisfies $-x =
\espalt{n}{\qprno V'(y_{n}\qprno)}$. By way of contradiction, assume there is some sequence (still labeled $n$) such that
$\lim_{n\uparrow\infty} y_n  = 0$.  Let $(M_n)_{n\in\nats}$ be such that
$\lim_{n\uparrow\infty} M_n= \infty$ and $\lim_{n\uparrow\infty} y_n M_n =
0$. For $n$ so large that $y_n < 1$, the strict convexity of $V$ gives
\begin{equation}\label{E:x_ub_1}
-x \leq V'(y_n M_n)\espalt{n}{\qprno 1_{\qprno\leq M_n}} +
\espalt{n}{\qprno V'(\qprno)1_{\qprno > M_n}}.
\end{equation}

The uniformly integrability
of $\qprno$ combined with $\espalt{n}{\qprno} = 1$,  $\lim_{z\downarrow 0}V'(z) = -\infty$ implies that
$\lim_{n\uparrow\infty} V'(y_nM_n)\espalt{n}{\qprno 1_{\qprno}\leq M_n} =
-\infty$.  From \cite[Corollary 4.2(ii)]{MR1865021} (note: part $(ii)$ therein does
not require $U(0) > 0$) there exists some  $\tilde{K}>0$
so that $z|V'(z)|\leq \tilde{K}V(z)$ for $z>0$.  Furthermore, since
$M_n\rightarrow\infty$, for any $\eps >0$, \eqref{E:dual_util_limit} (resp. \eqref{E:dual_util_limit_power}) and the
definitions of $V_\alpha$ (resp. $V_p$) imply that for large enough $n$:
\begin{equation*}
V(z)1_{z\geq M_n}\leq (1+\eps)\left(V_\alpha(z) + \frac{1}{\alpha}\right);\qquad
(\textrm{resp. } V(z)1_{z\geq M_n} \leq (1+\eps)V_p(z)).
\end{equation*}
In view of \eqref{E:relent_ub}, for some large enough $K$:
\begin{equation*}
\limsup_{n\uparrow\infty}\espalt{n}{\qprno
  V'(y_n\qprno)1_{\qprno > M_n}}\leq K;\qquad \left(\textrm{resp. }
\limsup_{n\uparrow\infty} \espalt{n}{\qprnt V'(y_n\qprnt)1_{\qprnt > M_n}}
\leq K\right).
\end{equation*}
Therefore, \eqref{E:x_ub_1} is contradicted if $y_n\rightarrow 0$, proving the result.
\end{proof}


\begin{proof}[Proof of Theorem \ref{T:uip_big_result}]

Let $\alpha > 0$, $U\in\Ua$ and $x\in\reals$. In view of Proposition
\ref{P:val_funct_no_claim}, one may choose $\eps > 0$ so that $\eps <
-u^n_U(x)$ for $n$ large. Recall the definition of $\alpha^n_U(\qprob^n)$ in
\eqref{E:penalty_functional} and the price $p^n_U(x,q;h^n)$ in
\eqref{E:util_indif_price_rep_pos}. Lemma \ref{L:Y_V_ubounds} with $u =
-u^n(x)$, $\eps = \eps$ and $Y = Z^{\qprob,n}$ implies there is a constant $\ol{C}(\eps,U)$ such that
\begin{equation*}
p^n_U(x,q_n;h^n) \leq \frac{x+\ol{C}(\eps,U)-u^n_U(x)}{q_n} +
\inf_{\qprob^n\in\tM^{n}}\left(\espalt{\qprob^n}{h^n} +
  \frac{1+\eps}{q_n\alpha}\relent{\qprob^n}{\prob^n}\right).
\end{equation*}
Similarly, from Lemma \ref{L:Y_V_lbounds} with $u = -u^n_U(x)$,
$\eps = \eps$ and $Y=Z^{\qprob,n}$ there exists constants $\ul{C}(\eps,U)$ and
$\ul{D}(\eps,U)$ such that
\begin{equation*}
p^n_U(x,q_n;h^n)\geq \frac{x -\ul{C}(\eps,U) + \ul{D}(\eps,U)\log(-u^n_U(x))}{q_n} +
\inf_{\qprob^n\in\tM^{n}}\left(\espalt{\qprob^n}{h^n} +
  \frac{1-\eps}{q_n\alpha}\relent{\qprob^n}{\prob^n}\right).
\end{equation*}
Consider the function:
\begin{equation}\label{E:f_delta_eq}
f(\delta,n) \dfn \inf_{\qprob^n\in\tM^{n}}\left(\espalt{\qprob^n}{h^n} +
  \delta\relent{\qprob^n}{\prob^n}\right),\quad \delta > 0.
\end{equation}
Clearly, $f$ is increasing with $\delta$.  Furthermore, Assumption
\ref{A:val_funct_limit} implies for some constant $K>0$ that $f(\delta,n) \leq
\|h\| + K \delta$. Let $0 < \delta < \gamma$.  For any $\qprob^n\in\tM^{n}$
\begin{equation*}
\begin{split}
\espalt{\qprob^n}{h^n} + \gamma \relent{\qprob^n}{\prob^n} &\leq \frac{\gamma}{\delta}\left(\espalt{\qprob^n}{h^n} +
  \delta \relent{\qprob^n}{\prob^n}\right) +\left(\frac{\gamma}{\delta}-1\right)\|h\|.
\end{split}
\end{equation*}
Thus,
\begin{equation}\label{E:exp_util_price_compare}
f(\gamma,n) - f(\delta,n) \leq \left(\frac{\gamma}{\delta}-1\right)\left(f(\delta,n)
  + \|h\|\right)\leq \left(\frac{\gamma}{\delta}-1\right)\left(2\|h\| + K\delta\right).
\end{equation}
Now, let $U_1,U_2\in\Ua$ and $x_1,x_2\in\reals$.  Choose $\eps > 0$ so that for all $n$
large enough $\eps \leq -u^n_{U_1}(x_1)\leq -U_1(x_1)$ and $\eps \leq -u^n_{U_2}(x_2)\leq -U_2(x_2)$. By the
above calculations, there is a constant $C(n,\eps)$ satisfying $C(n,\eps)/q_n
\rightarrow 0$ for any $q_n\rightarrow\infty$  such that
\begin{equation*}
\begin{split}
p^n_{U_1}(x_1,q_n;h^n) - p^n_{U_2}(x_2,q_n;h^n)&\leq \frac{C(n,\eps)}{q_n} +
f\left(\frac{1+\eps}{q_n\alpha}, n\right)  - f\left(\frac{1-\eps}{q_n\alpha}, n\right),\\
&\leq \frac{C(n,\eps)}{q_n} + \left(\frac{1+\eps}{1-\eps}-1\right)\left(2\|h\|
  + K\frac{1-\eps}{q_n\alpha}\right).
\end{split}
\end{equation*}
Therefore
\begin{equation}\label{E:difference_calc}
\limsup_{n\uparrow\infty}\left( p^n_{U_1}(x_1,q_n;h^n) - p^n_{U_2}(x_2,q_n;h^n)\right)\leq 2\|h\|\left(\frac{1+\eps}{1-\eps}-1\right).
\end{equation}
Since the left hand side does not depend upon $\eps$ taking $\eps\downarrow 0$ gives \eqref{E:price_come_together_alt} after noting that the roles of $U_1, U_2$ and $x_1,x_2$ may be switched.

\end{proof}


\begin{proof}[Proof of Theorem \ref{T:uip_main_result_alt}]

The proof is nearly identical to that of Theorem \ref{T:uip_big_result}. Let $\alpha > 0$ and $U\in\UUa$, $x\in\reals$. Proposition
\ref{P:val_funct_no_claim} implies one can find $\eps > 0$ so that $-u^n_U(x)\geq \eps$ for large $n$. Using the representation
for $p^n_U(x,q;h^n)$ in \eqref{E:util_indif_price_rep_pos} it follows from Lemma \ref{L:Y_V_ubounds} applied to $u=-u^n_U(x), Y=\qprnn$ that there is a
constant $\ol{C}(\eps,U)$ such that
\begin{equation*}
p^n_U(x,q_n;h^n) \leq \frac{x + \ol{C}(\eps,U)-u^n_U(x)}{q_n} +
\inf_{\qprob^n\in\tM^{n}}\left(\espalt{\qprob^n}{h^n} +
  \frac{1}{q_n\alpha}\relent{\qprob^n}{\prob^n}\right).
\end{equation*}
Similarly, Lemma \ref{L:Y_V_lbounds}
applied to $u=-u^n_U(x), Y=\qprnn$ yields the existence of constants
$\ul{C}(\eps,U)$, $\ul{D}(\eps,U)$ so that
\begin{equation*}
p^n_U(x,q_n;h^n)\geq \frac{x-\ul{C}(\eps,U) +\ul{D}(\eps,U)\log(-u^n_U(x))}{q_n} +
\inf_{\qprob^n\in\tM^{n}}\left(\espalt{\qprob^n}{h^n} +
  \frac{1}{q_n\alpha}\relent{\qprob^n}{\prob^n}\right).
\end{equation*}
Note that the variational problems in each of the above inequalities are the
same.  Now, let $U_1,U_2\in\UUa$ and $x_1,x_2\in\reals$. Choose $\eps > 0$ so
that $\eps \leq -u^n_{U_1}(x_1)\leq -U_1(x_1)$ and $\eps \leq
-u^n_{U_2}(x_2)\leq -U_2(x_2)$ for large $n$. By the above, there is a
constant $C(n,\eps)$ such that $\sup_n|C(n,\eps)| < \infty$ and
\begin{equation*}
q_n\left(p^n_{U_1}(x_1,q_n;h^n)-p^n_{U_2}(x_2,q_n;h^n)\right)\leq
C(n,\eps).
\end{equation*}
Thus, the result follows by first taking $\lim_{n\uparrow\infty}$ and then switching the roles of $U_1,U_2$, $x_1,x_2$.

\end{proof}


\begin{proof}[Proof of Example \ref{Ex:market_no_conv}]

By Theorem \ref{T:uip_big_result}, to show that $p^n_U(x,q_n;h)$ converges for all $\alpha >
0$, $U\in\Ua$ and $x\in\reals$ it suffices to consider the exponential utility
$U_\alpha$.  As in Remark
\ref{R:exp_util_px}, set  $p^n_\alpha \dfn
p^n_{U_\alpha}(q_n;h^n)$. \eqref{E:one_period_trinomial_px} gives $p^n_\alpha = -(1/(q_n\alpha))\log\left((1-e^{-q_n})e^{-\alpha q_n h} +
  e^{-q_n}\right)$, and hence $\lim_{n\uparrow\infty}p^n_\alpha =
\min\cbra{h,\alpha^{-1}}$.

It is now shown that $p^n_{\hat{U}}(x,q_n;h)$ cannot converge for
any $x\in\reals$.  To this end, set $p^n \dfn
p^n_{\hat{U}}(x,q_n;h)$.  From \eqref{E:one_period_trinomial_px}, $p^n$ satisfies
\begin{equation}\label{E:P_tri_2}
\hat{U}(x) = (1-e^{-q_n})\hat{U}(x+q_n(h-p^n)) + e^{-q_n}\hat{U}(x-q_np^n).
\end{equation}
Assume that $p = \lim_{n\uparrow\infty} p^n$ exists. Since it is clear $0
\leq p \leq h$, first assume that $0\leq p < h$.
Since $\hat{U}(\infty) = 0$, it follows that $\hat{U}(x) =
\lim_{n\uparrow\infty}e^{-q_n}\hat{U}(x-q_np^n)$. This implies
$\hat{U}(x-q_np^n) = \alpha_n \hat{U}(x) e^{q_n}$ where $\alpha_n\rightarrow 1$. Thus, recalling $p^n\rightarrow p$
\begin{equation}\label{E:L_U_inv_limit_4}
p= \lim_{n\uparrow\infty}
-\frac{1}{q_n}\hat{U}^{-1}\left(\alpha_n\hat{U}(x)e^{q_n}\right) =
\lim_{n\uparrow\infty}
-\frac{1}{q_n}\hat{U}^{-1}\left(-e^{\log(-\alpha_n\hat{U}(x)) + q_n}\right).
\end{equation}
Note that $(\log(-\alpha_n \hat{U}(x)))_{n\in\nats}$ forms a bounded sequence
since $\alpha_n\rightarrow 1$. Set $g(z)\dfn U^{-1}(-e^{z})$. A straightforward calculation shows that $g'(z)
= U(g(z))/U'(g(z))$.  Since $g(z)\rightarrow -\infty$ as $z\rightarrow\infty$
it follows by \lopital's rule and the fact that $1/K_U \leq \alpha_U(x)\leq
K_U$ that
\begin{equation*}
-K_U \leq \liminf_{z\uparrow\infty} g'(z) \leq \limsup_{z\uparrow \infty} g'(z) \leq -\frac{1}{K_U}.
\end{equation*}
Thus, for large enough $n$
\begin{equation*}
\begin{split}
\frac{1}{q_n}\left|U^{-1}(-e^{\log(-\alpha_n\hat{U}(x)) + q_n}) - U^{-1}(-e^{q_n})\right|
&\leq 2K_U\frac{|\log(-\alpha_n\hat{U}(x))|}{q_n}.
\end{split}
\end{equation*}
Therefore, in view of \eqref{E:L_U_inv_limit_4} it follows that $p =
\lim_{n\uparrow\infty} -(1/q_n)\hat{U}^{-1}\left(-e^{q_n}\right)$, but this violates the assumption on $q_n$ and hence $p^n$ cannot converge
to $p < h$. Next, assume $p = h$ and let $\eps > 0$ be small enough so that $h - \eps > 0$. For large enough $n$, $p^n \geq h - \eps$. The negativity of $\hat{U}(x)$ and \eqref{E:P_tri_2} imply
\begin{equation*}
-\hat{U}(x) \geq e^{-q_n}(-\hat{U}(x-q_np^n)) \geq e^{-q_n}\left(-\hat{U}(x-q_n(h-\eps))\right).
\end{equation*}
This gives
\begin{equation*}
\begin{split}
0 &\geq -1 + \limsup_{n\uparrow\infty} \frac{1}{q_n}\log\left(-\hat{U}(x-q_n(h
  -\eps))\right) \geq -1 + \ul{\alpha}(h-\eps),
\end{split}
\end{equation*}
where the last inequality follows by \eqref{E:u_no_conv_facts}. Taking
$\eps\downarrow 0$ gives that $\ul{\alpha} \leq 1/h$ but, this violates how
$h$ was constructed.  Therefore, $p\neq h$ and the proof is complete.
\end{proof}


\begin{proof}[Proof of Example \ref{Ex:total_quant_diverge}]

Set $p^n_\alpha \dfn p^n_{U_\alpha}(x,n^2;h^n)$. Specifying
\eqref{E:one_period_trinomial_px} for the given parameter values gives
\begin{equation}\label{E:P_tri_23}
n^2 p^n_\alpha = -\frac{1}{n^2\alpha}\log\left((1-e^{-n})e^{-\alpha n^2} +
  e^{-n}\right) = \frac{n}{\alpha} + R(n),
\end{equation}
where $\lim_{n\uparrow\infty}R(n) = 0$.  For the given function
$U$ and $x\in\reals$, set $p^n \dfn p^n_U(x,n^2;h)$. Using
\eqref{E:one_period_trinomial_px} again:
\begin{equation}\label{E:P_tri_24}
U(x) = (1-e^{-n})U(x+n^2(1-p^n)) + e^{-n}U(x-n^2p^n).
\end{equation}
Assume that $\liminf_{n\uparrow\infty}n^2(p^n - p^n_\alpha) < \infty$ and choose
a subsequence $\cbra{n_k}_{k\in\nats}$ and $K >0$ so that $n_k^2(p^{n_k}-p^{n_k}_\alpha) \leq K$
for all $k$.  By the monotonicity of $U$ it follows from \eqref{E:P_tri_24}
that
\begin{equation}\label{E:P_tri_25}
U(x) \geq (1-e^{-n_k})U(x + n_k^2 - n_k^2 p^{n_k}_\alpha  -K) + e^{-n_k}U(x -
n_k^2p^{n_k}_\alpha - K).
\end{equation}
From \eqref{E:P_tri_23} and $U(\infty) = 0$ it follows that
\begin{equation*}
\lim_{k\uparrow\infty} (1-e^{-n_k})U(x + n_k^2 - n_k^2 p^{n_k}_\alpha  -K) =
0.
\end{equation*}
By construction of $U$ and \eqref{E:P_tri_23} again
\begin{equation*}
\lim_{k\uparrow\infty} e^{-n_k}U(x-n_k^2 p^{n_k}_\alpha - K) =
\lim_{k\uparrow\infty} -\frac{1}{\alpha}e^{-n_k}\frac{e^{-\alpha(x-K) + n_k +
  \alpha R(n)}}{n_k/\alpha - (x-K) + R(n)} = 0.
\end{equation*}
Therefore, \eqref{E:P_tri_25} implies $U(x) \geq 0$, a contradiction. Thus, $\lim_{n\uparrow\infty} n^2(p^n-p^n_\alpha) = \infty$.

\end{proof}


\begin{proof}[Proof of Proposition \ref{P:uip_big_result_power}]

The proof is very similar to that of Theorem \ref{T:uip_big_result}.  Namely,
let $p > 1$, $l>0$, $U\in\Up$ and $x\in\reals$.  Proposition
\ref{P:val_funct_no_claim} implies for $\eps>0$ small enough, $\eps < -u^n_U(x)$ for large $n$. Lemma
\ref{L:Y_V_ubounds} with $u=-u^n_U(x)$, $\eps = \eps$ and $Y = \qprnn$ yields a constant $\ol{C}(\eps,U)$ so that for all position sizes $q$ (and
not just the $q_n$ of the Proposition)
\begin{equation*}
p^n_U(x,q;h^n) \leq \frac{x+\ol{C}(\eps,U)}{q} +
\inf_{\qprob^n\in\hM^{n}}\left(\espalt{\qprob^n}{h^n} + \frac{1}{q}\left(l(-u^n_U(x)+\eps)\right)^{1/p}\left(1+\eps\right)^{1/\gamma}\espalt{n}{(\qprnn)^\gamma}^{1/\gamma}\right).
\end{equation*}
Similarly, from Lemma \ref{L:Y_V_lbounds} with $u = -u^n_U(x)$,
$\eps = \eps$ and $Y=Z^{\qprob,n}$ there exists a constant $\ul{C}(\eps,U)$
such that
\begin{equation*}
p^n_U(x,q;h^n) \geq \frac{x-\ul{C}(\eps,U)}{q} +
\inf_{\qprob^n\in\hM^{n}}\left(\espalt{\qprob^n}{h^n} + \frac{1}{q}\left(l(-u^n_U(x)-\eps/2)\right)^{1/p}\left((1-\eps)\right)^{1/\gamma}\espalt{n}{(\qprnn)^\gamma}^{1/\gamma}\right).
\end{equation*}
Now, consider the function:
\begin{equation}\label{E:hf_delta_eq}
\hat{f}(\delta,n) \dfn \inf_{\qprob^n\in\hM^{n}}\left(\espalt{\qprob^n}{h^n} +
  \delta\espalt{n}{\left(\qprnn\right)^\gamma}^{1/\gamma}\right),\quad \delta > 0.
\end{equation}
$\hat{f}$ is increasing with $\delta$ and that Assumption
\ref{A:val_funct_limit_power} implies the existence of a constant $K>0$ so
that $\hat{f}(\delta,n) \leq \|h\| + K \delta$. Let $0 < \delta < \gamma$.  For any $\qprob^n\in\hM^{n}$
\begin{equation*}
\begin{split}
\espalt{\qprob^n}{h^n} + \gamma \relent{\qprob^n}{\prob^n} &\leq \frac{\gamma}{\delta}\left(\espalt{\qprob^n}{h^n} +
  \delta \espalt{n}{\left(\qprnn\right)^\gamma}^{1/\gamma}\right) +\left(\frac{\gamma}{\delta}-1\right)\|h\|.
\end{split}
\end{equation*}
Thus,
\begin{equation}\label{E:exp_util_price_compare_power}
\hat{f}(\gamma,n) - \hat{f}(\delta,n) \leq \left(\frac{\gamma}{\delta}-1\right)\left(\hat{f}(\delta,n) + \|h\|\right)\leq \left(\frac{\gamma}{\delta}-1\right)\left(K\delta + 2\|h\|\right).
\end{equation}
Note that for any $x\in\reals$, $U\in\Up$, since $u^n_U(x) \geq U(x)$ and $\limsup_{n\uparrow\infty} u^n_U(x) < 0$ there exists some $M >0$ so that $1/M \leq -u^n_U(x) \leq M$ for all $n$ large enough. Now, let $U_1,U_2\in\Ua$ and $x_1,x_2\in\reals$ and consider $q_n\uparrow\infty$.  Choose $M > 0, \eps > 0$ so that for all $n$
large enough $\eps < 1/M \leq -u^n_{U_i}(x_i)\leq M; i=1,2$. By the above calculations
\begin{equation}\label{E:power_t1}
\begin{split}
&p^n_{U_1}(x_1,q_n(-u^n_{U_1}(x_1))^{1/p};h^n) - p^n_{U_2}(x_2,q_n(-u^n_{U_2}(x_2))^{1/p};h^n)\\
&\qquad\qquad\leq \frac{C^+(\eps,n)-C^-(\eps,n)}{q_n} + \hat{f}\left(\frac{\delta^+(\eps,n)}{q_n},n\right) - \hat{f}\left(\frac{\delta^-(\eps,n)}{q_n},n\right),
\end{split}
\end{equation}
where
\begin{equation*}
\begin{split}
C^+(\eps,n)&\dfn \frac{x_1+\ol{C}(\eps,U_1)}{(-u^n_{U_1}(x_1))^{1/p}};\qquad C^-(\eps,n)\dfn
\frac{x_2+\ul{C}(\eps,U_2)}{(-u^n_{U_2}(x_2))^{1/p}},\\
\delta^+(\eps,n)&\dfn \left(1-\frac{\eps}{u^n_{U_1}(x_1)}\right)^{1/p}l^{1/p}(1+\eps)^{1/\gamma};\qquad \delta^-(\eps,n)\dfn \left(1+\frac{\eps/2}{u^n_{U_1}(x_1)}\right)^{1/p}l^{1/p}(1-\eps)^{1/\gamma}.
\end{split}
\end{equation*}
Since $C^\pm(\eps,n)/q_n\rightarrow 0$ as $n\uparrow\infty$ for all $\eps >0$ they may be disregarded. Also, note that $\delta^+(\eps,n) > \delta^-(\eps,n)$ and 
\begin{equation*}
\frac{\delta^+(\eps,n)}{\delta^-(\eps,n)} \leq \frac{(1+\eps M)^{1/p}(1+\eps)^{1/\gamma}}{(1-\eps M/2)^{1/p}(1-\eps)^{1/\gamma}};\qquad \lim_{n\uparrow\infty} \frac{\delta^{-}(\eps,n)}{q_n} = 0.
\end{equation*}
It thus follows from \eqref{E:power_t1} and \eqref{E:exp_util_price_compare_power} that for all $\eps > 0$
\begin{equation}\label{E:power_t1}
\begin{split}
&\limsup_{n\uparrow\infty} \left(p^n_{U_1}(x_1,q_n(-lu^n_{U_1}(x_1))^{1/p};h^n) - p^n_{U_2}(x_2,q_n(-lu^n_{U_2}(x_2))^{1/p};h^n)\right)\\
&\qquad\qquad\leq 2\|h\|\left(\frac{(1+\eps M)^{1/p}(1+\eps)^{1/\gamma}}{(1-\eps M/2)^{1/p}(1-\eps)^{1/\gamma}}-1\right).
\end{split}
\end{equation}
Taking $\eps\downarrow 0$ gives the result after noting that the roles of $U_1, U_2$ and $x_1,x_2$ may be switched.

\end{proof}


\section{Proofs from Section \ref{S:basis_risk}}\label{S:more_proofs_2}

\subsection{Preliminaries} A key fact used in all the proofs below is that under Assumption
\ref{A:basis_risk}, the value function for the exponential utility $U_\alpha$
takes the form  \cite[Proposition
3.3]{MR2094149}:
\begin{equation}\label{E:basis_risk_model_vf}
u^n_{U_{\alpha}}(0,q_n;h) = -\frac{1}{\alpha}\espalt{\prob}{Z(\rho_n)\xpn{-(1-\rho_n^2)\left(\alpha
      q_n h(Y_T)+\frac{1}{2}\!\int_0^T\!\lambda(Y_t)^2dt\right)}}^{\frac{1}{1-\rho_n^2}},
\end{equation}
where $Z(\rho_n)$ is given in \eqref{E:z_rho_def}. Recall that $\Lambda = (1/2)\int_0^T\lambda(Y_t)^2dt$  and note that
$Z(\rho_n)e^{-(1-\rho^2_n)\Lambda} = Z^{\rho_n}e^{-(1-\rho_n)\Lambda}$ where
$Z=Z(1)$. In accordance with \eqref{E:opt_qn} set $\gamma_n =
\alpha(1-\rho_n^2)q_n$. Using \eqref{E:basis_risk_model_vf} and the definition of $p^n_{U_{\alpha}}(q;h^n)$ in
\eqref{E:uip_price_def}
\begin{equation}\label{E:br_exp_util_indif_px}
p^n_{U_{\alpha}}\left(q_n;h\right) =
-\frac{1}{q_n\alpha}\log\left(\frac{u^n_{U_\alpha}(0,q_n;h)}{u^n_{U_\alpha}(0)}\right)
= -\frac{1}{\gamma_n}\log\left(\frac{\espalt{}{Z^{\rho_n}e^{-(1-\rho_n)\Lambda - \gamma_n
      h(Y_T)}}}{\espalt{}{Z^{\rho_n}e^{-(1-\rho_n)\Lambda}}}\right),
\end{equation}

Therefore, the proofs of Propositions
\ref{P:uip_price_br} and \ref{P:uip_error_br} rely heavily on the analysis of the function
\begin{equation}\label{E:f_rho_gamma}
f(\rho,\gamma) \dfn  \frac{\espalt{}{Z^{\rho}e^{-(1-\rho)\Lambda - \gamma
      h(Y_T)}}}{\espalt{}{Z^{\rho}e^{-(1-\rho)\Lambda}}};\quad
\rho,\gamma\in\reals.
\end{equation}
The bounded-ness of $\Lambda$, $h(Y_T)$ combined with the fact that $Z$ has
exponential moments of all orders, imply that $f$ is smooth
in $(\rho,\gamma)$ and that derivatives may be computed by pulling the
differentiation operator through the expected value operator. Furthermore, for
the function $g$ from \eqref{E:g_rho_gamma}, 
$g(\rho,\gamma) = -\partial_\gamma \log(f(\rho,\gamma))$.

\subsection{Proofs} The following Lemma is used throughout and is a basic
result on Esscher transformations. Recall from Remark \ref{R:fast_case} that
$\essinf{\prob}{h(Y_T)} = \inf_{y\in E}h(y)$ and (similarly)
$\esssup{\prob}{h(Y_T)} = \sup_{y\in E}h(y)$. 

\begin{lemma}\label{L:h_funct_facts}
Let Assumption \ref{A:basis_risk} hold.  For $\rho,\gamma\in\reals$ let
$g(\rho,\gamma)$ be as in \eqref{E:g_rho_gamma}. Then
\begin{enumerate}[i)]
\item For $\rho$ fixed, $g$ is strictly decreasing in $\gamma$ with
  $\lim_{\gamma\downarrow -\infty}g(\rho,\gamma) = \sup_{y\in E}h(y)$ and
  $\lim_{\gamma\uparrow\infty}g(\rho,\gamma) = \inf_{y\in E}h(y)$.
\item For $p\in (\inf_{y\in E}h(y), \sup_{y\in E}h(y))$ and $\rho\in\reals$, there
  exists a unique $\gamma = \gamma(\rho)$ such that $p =
  g(\rho,\gamma(\rho))$.  The map $\rho\mapsto \gamma(\rho)$ is $C^1$ on
  $\reals$.
\end{enumerate}
\end{lemma}


\begin{proof}[Proof of Lemma \ref{L:h_funct_facts}]
Calculation shows
\begin{equation*}
\partial_{\gamma}g(\rho,\gamma) = -\varalt{\tilde{\prob}}{h(Y_T)};\qquad
\frac{d\tilde{\prob}}{d\prob} \dfn \frac{Z^{\rho}e^{-(1-\rho)\Lambda -\gamma
    h(Y_T)}}{\espalt{}{Z^{\rho}e^{-(1-\rho)\Lambda -\gamma h(Y_T)}}}.
\end{equation*}
so that $g$ is strictly decreasing. Set $\ul{h} \dfn \essinf{\prob}{h(Y_T)} =
\inf_{y\in E}h(y)$. Clearly, $g(\rho,\gamma) \geq
\ul{h}$. Now, let $m > 0$ be such that $\probalt{h(Y_T) -\ul{h}< m}
>0$ and $\probalt{h(Y_T)-\ul{h} \geq m} > 0$. For $\gamma >0$ and $K > 0$
large enough
\begin{equation*}
\begin{split}
g(\rho,\gamma) &= \ul{h} + \frac{\espalt{}{(h(Y_T)-\ul{h})Z^{\rho}e^{-(1-\rho)\Lambda-\gamma
      (h(Y_T)-(\ul{h}+m))}}}{\espalt{}{Z^{\rho}e^{-(1-\rho)\Lambda-\gamma(h(Y_T)-(\ul{h}+m))}}},\\
&\leq \ul{h} + m +
\frac{K}{\espalt{}{Z^{\rho}e^{-(1-\rho)\Lambda-\gamma(h(Y_T)-(\ul{h}+m))}1_{h(Y_T)
    < \ul{h}+m}}}.
\end{split}
\end{equation*}
Fatou's Lemma now yields $\limsup_{\gamma\uparrow\infty}
g(\rho,\gamma)\leq \ul{h}+
m$. Taking $m\downarrow 0$ gives the result. The result for
$\lim_{\gamma\downarrow -\infty} g(\rho,\gamma)$ follows by a similar
argument.

As for $ii)$, part $i)$ clearly gives, for each $\rho\in\reals$ and each $p\in
I(h)$, a unique $\gamma(\rho)$ such that $p = g(\rho,\gamma(\rho))$. The
result now follows by the Implicit Function Theorem \cite[Theorem
9.28]{MR0385023}, since $g$ is smooth in $\rho,\gamma$ and, as was shown in
part $i)$, $\partial_\gamma g(\rho,\gamma) \neq 0$.
\end{proof}


\begin{proof}[Proof of Proposition \ref{P:uip_price_br}]

Recall $p^n_{U_\alpha}$ from \eqref{E:br_exp_util_indif_px} and $f$ from \eqref{E:f_rho_gamma}. Case $(i)$ is handled first. Here, Taylor's formula and $f(\rho_n,0) =
1$ yield
\begin{equation*}
\begin{split}
f(\rho_n,\gamma_n) &=  1 +  \gamma_n\partial_\gamma f(\rho_n,0) +
(1/2)\gamma_n^2\partial^2_{\gamma\gamma}f(\rho_n,\xi_n);\qquad 0\leq \xi_n\leq \gamma_n.
\end{split}
\end{equation*}
Since for all $(\rho,\gamma)$,  $|\partial_\gamma f(\rho,\gamma)| \leq \sup_{y\in E}|h(y)|$ and
$|\partial^2_{\gamma\gamma}f(\rho,\gamma)|\leq \sup_{y\in E}|h(y)|^2$, the
approximation $\log(1 + x) \approx x$ for small $x$ implies that $\lim_{n\uparrow\infty}
p^n_{U_\alpha}(q_n;h) =  -\partial_\gamma f(1,0) = \espalt{}{Zh(Y_T)}$. The
results for case $(ii)$ follow immediately from the continuity of $f$ and $f(1,\gamma) =
\espalt{\qprob}{\xpn{-\gamma h(Y_T)}}$. As for case $(iii)$, set $\ul{h} = \essinf{\prob}{h(Y_T)}$.  Clearly
$\liminf_{n\uparrow\infty}p^n_{U_\alpha}(q_n;h)\geq \ul{h}$.  Now, let $m >
\ul{h}$, $A_m = \cbra{h(Y_T) < m}$ and note that $\probalt{A_m} > 0$.  Then
\begin{equation*}
\espalt{}{Z^{\rho_n}e^{(1-\rho_n)\Lambda - \gamma_n
    h(Y_T)}} \geq
e^{-\gamma_n
  m}\espalt{}{Z^{\rho_n}e^{-(1-\rho_n)\Lambda}1_{A_m}},
\end{equation*}
so that, using \eqref{E:f_rho_gamma} and \eqref{E:br_exp_util_indif_px}, $\limsup_{n\uparrow\infty}p^n_{U_\alpha}(q_m;h)\leq m$.  The result
follows taking $m\downarrow \ul{h}$.
\end{proof}



\begin{proof}[Proof of Proposition \ref{P:br_opt_q}]

Recall $g$ from \eqref{E:g_rho_gamma}. \cite[Theorem 3.1]{MR2212897} gives that the optimal $q_n$ satisfies the
first order conditions
\begin{equation*}
p_n = -\frac{1}{\alpha}\frac{\partial_q
  u^n_{U\alpha}(0,q_n;h)}{u^n_{U_\alpha}(0,q_n;h)} = g(\rho_n,\alpha
q_n(1-\rho^2_n)) = g(\rho_n,\gamma_n),
\end{equation*}
in view of \eqref{E:opt_qn}. That such a $\gamma_n$ exists and is unique follows from Lemma
\ref{L:h_funct_facts} part $i)$. Now, let $\rho_n\rightarrow 1$ and note that
\begin{equation*}
\frac{p_n-\hat{p}}{1-\rho^2_n} = \frac{g(\rho_n,\gamma_n) -
  g(1,0)}{1-\rho^2_n}.
\end{equation*}
Assume that $\sup_n |p_n-\hat{p}|/(1-\rho^2_n) < \infty$. Then Lemma
\ref{L:h_funct_facts} implies 
$\gamma_n\rightarrow 0$ and hence by the first order Taylor approximation
(higher orders may be ignored since $\rho_n\rightarrow 1, \gamma_n\rightarrow 0$)
\begin{equation*}
\frac{p_n-\hat{p}}{1-\rho^2_n} = \frac{g(\rho_n,\gamma_n) - g(1,0)}{1-\rho^2_n} \approx \frac{-1}{1+\rho_n}\partial_\rho g(1,0)
+ \frac{\gamma_n}{1-\rho^2_n}\partial_\gamma(1,0).
\end{equation*}
The equivalence in \eqref{E:q_n_p_n_lim} readily follow. Lastly, assume that $p_n\rightarrow p \neq \hat{p}$. Then by continuity
$\gamma_n\rightarrow \gamma$ where $\gamma$ satisfies $p =
g(1,\gamma)$. Thus, by Lemma \ref{L:h_funct_facts}, $\gamma \neq 0$ if and
only if $p\neq \hat{p}$. 

\end{proof}


\begin{proof}[Proof of Proposition \ref{P:uip_error_br}]

Here, the monetary error takes the form
\begin{equation*}
\textrm{ME}_n \dfn q_n\left|p^n_{U_\alpha}(q_n;h) - p_\alpha\right| =
\frac{\gamma_n}{\alpha(1-\rho^2_n)}\left|-\frac{1}{\gamma_n}\log(f(\rho_n,\gamma_n))
  - p_\alpha\right|.
\end{equation*}
To estimate the error it is now necessary to go out two terms in the Taylor
expansion for $\log(f(\rho_n,\gamma_n))$ around $\log(f(1,\gamma))$ (higher order terms may be
ignored as discussed below). Note that for all $\gamma\geq 0$, $f(1,\gamma) >0$ and the first and
second order partial derivatives of $f$ exist and are finite. Thus, since
$(1-\rho_n)/(1-\rho_n^2) \leq 1$, terms in the Taylor expansion which involve
partials with respect to $\rho$, even when divided by $1-\rho^2_n$, remain
bounded as $n\uparrow\infty$. Hence, the only terms which will affect the finiteness of the monetary
error involve $\log(f(1,\gamma))$, 
$\partial_\gamma \log(f(1,\gamma))$ and
$\partial^2_{\gamma\gamma}\log(f(1,\gamma)$ and it suffices to
consider the approximation
\begin{equation*}
\begin{split}
\log(f(\rho_n,\gamma_n)) &\approx \log(f(1,\gamma)) + (\gamma_n-\gamma)\frac{\partial_\gamma f}{f}(1,\gamma) +
\frac{1}{2}(\gamma_n-\gamma)^2\left(\frac{\partial^2_{\gamma\gamma}f}{f} -
  \left(\frac{\partial_\gamma f}{f}\right)^2\right)(1,\gamma).
\end{split}
\end{equation*}
Note that $\partial^2_{\gamma\gamma}f / f - (\partial_\gamma f/f)^2
= -\partial_\gamma g$ where $g$ is from \eqref{E:g_rho_gamma}.  By Lemma
\ref{L:h_funct_facts}, $\partial_\gamma g < 0$. Case $(i)$ is handled first : here $p_\alpha=-\partial_\gamma \log(f)(1,0) =
\espalt{}{Zh(Y_T)}$.  Using the above expansion, and the fact that $f(1,0) = 1$
\begin{equation*}
\textrm{ME}_n \approx
\frac{\gamma^2_n}{2\alpha(1-\rho_n^2)}\left|(\partial^2_{\gamma\gamma}f - (\partial_\gamma f)^2)(1,0)\right|,
\end{equation*}
and hence the result follows.  Now, consider case $(ii)$ : here $p_\alpha =
-(1/\gamma)\log(f(1,\gamma))$.  Using the above expansion it follows that
\begin{equation*}
\textrm{ME}_n \approx
\frac{|\gamma_n-\gamma|}{\alpha(1-\rho_n^2)}\left|\frac{\partial_\gamma
    f}{f}(1,\gamma) + \frac{1}{2}(\gamma_n-\gamma)\left(\frac{\partial^2_{\gamma\gamma} f}{f} -
    \left(\frac{\partial_\gamma
        f}{f}\right)^2\right)(1,\gamma)-\frac{1}{\gamma}\log f(1,\gamma)\right|,
\end{equation*}
proving the result for case $(ii)$.  Now, let $p\in I(h)$ and assume
$\gamma_n$ is chosen optimally from Proposition \ref{P:br_opt_q} so that $p =
g(\rho_n,\gamma_n)$.  If $p=\hat{p}$ then by \eqref{E:q_n_p_n_lim}, since $p_n =
\hat{p}$ for all $n$ it holds that $\sup_n |q_n| < \infty$ and hence the
monetary error is trivially finite.  If $p\neq \hat{p}$, by Proposition
\ref{P:br_opt_q} again, $\gamma_n \rightarrow \gamma \neq 0$ where
$p=g(1,\gamma)$. Therefore, $\gamma_n$ is in regime $(ii)$ of
\eqref{E:opt_qn} and hence the monetary error is finite if and only if
$\sup_{n}|\gamma_n-\gamma|/(1-\rho^2_n) < \infty$. Using the notation in part $ii)$ of Lemma
\ref{L:h_funct_facts}, $\gamma_n
= \gamma(\rho_n)$ and $\gamma = \gamma(1)$.  Since the map $\rho \mapsto
\gamma(\rho)$ is $C^1$ on $\reals$, for any $\eps > 0$ by taking $n$ large enough
\begin{equation*}
\frac{|\gamma_n - \gamma|}{1-\rho^2_n} =
\frac{|\gamma(\rho_n)-\gamma(1)|}{1-\rho^2_n} =
\frac{1}{1+\rho_n}\frac{1}{1-\rho_n}\left|\int_{\rho_n}^1
  \gamma'(\tau)d\tau\right| \leq \frac{|\gamma'(1)| + \eps}{2},
\end{equation*}
and hence the monetary error is bounded. 
\end{proof}


\begin{proof}[Proof of Proposition \ref{P:uip_mu_facts}]

As shown in \cite{MR2094149}, the dual optimal element
$\qprob^{n,q_n}$ is constructed using the Martingale Representation
Theorem. Specifically, there exists an $\F^W$ adapted process $\theta^n$
such that \eqref{E:mart_rep_opt_strat} holds. For this $\theta_n$, define $\qprob^{n,q_n}$ by
\begin{equation*}
\frac{d\qprob^{n,q_n}}{d\prob} \dfn
\mathcal{E}\left(\int_0^\cdot\left(-\rho_n\lambda(Y_t) + \theta^n_t\right)dW_t -
  \int_0^\cdot\left(\sqrt{1-\rho^2_n}\lambda(Y_T) - \frac{\rho_n}{\sqrt{1-\rho^2_n}}\theta^n_t\right)dB_t\right)_T.
\end{equation*}
It is easy to check that $\qprob^{n,q_n}\in\M^n$ and a 
calculation shows that $\relent{\qprob^{n,q_n}}{\prob} < \infty$. Furthermore,
$\qprob^{n,q_n}$ solves the dual problem. This latter fact follows by
considering the (potentially non-admissible) trading strategy $\pi_t \dfn
(1/\alpha\sigma(Y_t))(\lambda(Y_t) + (\rho_n)/(1-\rho^2_n)\theta^n_t)$ and
showing that the corresponding wealth process $X^\pi$ and
$d\qprob^{n,q_n}/d\prob$ satisfy the first order conditions for optimality. Then, from
\cite[Theorem 2.1]{MR1891731} it follows that $\int \pi_udS_u/S_u$ is a $\qprob^{n,q_n}$
martingale and hence $\qprob^{n,q_n}$ is dual
optimal. A straightforward calculation using \eqref{E:hat_qn_def} gives
\begin{equation*}
\frac{d\qprob^{n,q_n}_W}{d\hat{\qprob}^n_W} = \frac{e^{
    -(1-\rho^2_n)\Lambda-\gamma_n
    h(Y_T)}}{\espalt{}{Z(\rho_n)e^{-(1-\rho^2_n)\Lambda-\gamma_n h(Y_T)}}}.
\end{equation*}
Therefore, using \eqref{E:br_exp_util_indif_px}, \eqref{E:g_rho_gamma} and
that $p = g(\rho_n,\gamma_n)$ for $g$ from \eqref{E:g_rho_gamma}:
\begin{equation*}
\begin{split}
\relent{\qprob^{n,q_n(p)}_W}{\hat{\qprob}^n_W} &= \gamma_n(p^n_{U_\alpha}(q_n;h)-p) - (1-\rho^2_n)\frac{\espalt{}{\Lambda Z(\rho_n)e^{
      -(1-\rho^2_n)\Lambda - \gamma_n h(Y_T)}}}{\espalt{}{Z(\rho_n)e^{ -
      (1-\rho^2_n)\Lambda-\gamma_n h(Y_T)}}}\\
&\qquad\qquad  -\log\left(\espalt{}{Z(\rho_n)e^{-(1-\rho^2_n)\Lambda}}\right),
\end{split}
\end{equation*}
and so the result follows by Proposition \ref{P:uip_price_br} since
$\gamma_n\rightarrow\gamma\neq 0$.

\end{proof}


\appendix

\section{Supporting Lemmas}\label{S:V_rv_lemmas}

\begin{remark}\label{R:V_rv_notation}Throughout this section, $\probtriple$ represents a generic probability
space and all expectations, unless explicitly stated otherwise, are with respect to
$\prob$. Inequalities regarding random variables are assumed to hold $\prob$
almost surely.
\end{remark}


\begin{lemma}\label{L:dual_class_equiv}
Let $Y\geq 0$. The following three statements are equivalent:
\begin{enumerate}[1)]
\item $\espalt{}{V(yY)} < \infty$ for all $\alpha > 0$, $U\in\Ua$ and $y>0$.
\item $\espalt{}{V(yY)} < \infty$ for some $\alpha > 0$, $U\in\Ua$ and $y>0$.
\item $\espalt{}{Y\log Y} < \infty$.
\end{enumerate}
Furthermore, let $p>1$ and set $\gamma = p/(p-1)$. Then the following three statements are also equivalent:
\begin{enumerate}[A)]
\item $\espalt{}{V(yY)} < \infty$ for all $l>0$, $U\in\Up$ and $y>0$.
\item $\espalt{}{V(yY)} < \infty$ for some $l > 0$, $U\in\Up$ $y>0$.
\item $\espalt{}{Y^{\gamma}} < \infty$.
\end{enumerate}
\end{lemma}


\begin{proof}[Proof of Lemma \ref{L:dual_class_equiv}]

Let $\alpha > 0$ and $U\in\Ua$.  In view of \eqref{E:dual_util_limit}, for any
$\eps > 0$ there is some constant $M = M(\eps,U)
> 0$ such that on $z\geq M$, $((1-\eps)/\alpha)z(\log(z)-1)\leq V(z) \leq
((1+\eps)/\alpha)z(\log(z)-1)$. Since both $|V(z)|$ and $|z\log(z)-1|$ are
bounded on $[0,M]$, there is some $C = C(\eps,M)>0$ so that
\begin{equation}\label{eq: dual_funct_ubounds}
\begin{split}
-C + \frac{1-\eps}{\alpha}z\log(z)\leq V(z)\leq  C + \frac{1+\eps}{\alpha}z\log(z).
\end{split}
\end{equation}
The equivalences $1)\Leftrightarrow 2)\Leftrightarrow 3)$ now readily follow.
Similarly, in view of \eqref{E:dual_util_limit_power} for any $\eps > 0$ there
is some constant $M = M(\eps,U)$ so that on $z\geq M$, $(1-\eps)(1/\gamma)\hat{l} z^\gamma
\leq V(z) \leq (1+\eps)\hat{l} z^\gamma$. Again, since
$|V(z)|$ and $z^\gamma$ are bounded on $[0,M]$ there is a constant $C =
C(\eps,U)$ so that
\begin{equation}\label{eq: dual_funct_bounds_power}
-C + (1-\eps)\hat{l} z^\gamma \leq
V(z) \leq C + (1+\eps)\hat{l}z^\gamma.
\end{equation}
The equivalences $A)\Leftrightarrow B)\Leftrightarrow C)$ readily follow.

\end{proof}


\begin{lemma}\label{L:some_Y_V_facts}

Let $\alpha > 0$, $p>1$, $l>0$.  Let $U\in\Ua$ or $U\in\Up$. Let $Y\geq 0$ be such that
$\espalt{}{Y} = 1$ and such that $\espalt{}{V(Y)} < \infty$. Then the map
$y\mapsto \espalt{}{V(yY)}$ is differentiable with derivative
$\espalt{}{YV'(yY)}$.  Furthermore, for any $x\in\reals$ there exists a unique
$y$ such that $\espalt{}{Y V'(yY)} = x$.
\end{lemma}


\begin{proof}

Consider the function
\begin{equation*}
f(\eps,z) \dfn \frac{V((y+\eps)z) - V(yz)}{\eps} - \frac{V(yz)}{y};\qquad \eps >
0,z\geq 0.
\end{equation*}
Note that $f(\eps,0)=0$ and, as $V'$ is strictly increasing, $\partial_z
f(\eps,z)\geq 0$.  The convexity of $V$ implies $f(\eps,Y)\leq V((1+y)Y) - V(yY) - V(yY)/y$. That $\partial_y \espalt{}{V(yY)} = \espalt{}{YV'(yY)}$
now follows by applying the dominated convergence theorem to
$f(\eps,Y)$, since Lemma \ref{L:dual_class_equiv} implies $\espalt{}{V(yY)} <
\infty$ for all $y > 0$.

Now, consider the map $g(y) \dfn \espalt{}{YV'(yY)}$. The
strict convexity of $V$ implies $g$ is strictly increasing. Since
$\lim_{z\downarrow} zV'(z) = 0$ (because $U$ is bounded from
above), there is some constant $C > 0$ such that $YV'(yY)>-C$ for $y>1$.
Thus, since $\lim_{z\uparrow\infty} V'(z) = \infty$ (because of the Inada conditions) it follows by Fatou's Lemma that $\lim_{y\uparrow\infty}
g(y)=\infty$.  For the limit as $y\downarrow 0$, assume $y < 1$. Denote by $\hat{y}$
by the unique number such that $V'(\hat{y}) = 0$. Clearly
\begin{equation*}
g(y) = \espalt{}{YV'(yY)1_{yY\leq \hat{y}}} +
\espalt{}{YV'(yY)1_{yY > \hat{y}}}.
\end{equation*}
Fatou's Lemma and $\lim_{z\downarrow 0}V'(z) = -\infty$ imply that
$\lim_{y\downarrow 0} \espalt{}{YV'(yY)1_{yY\leq \hat{y}}} = -\infty$. As for
the second term, since $y<1$, $V'(yY)\leq V'(Y)$.  Thus, using \cite[Corollary
4.2 (ii)]{MR1865021} (note : part $(ii)$ therein does not require $U(0) > 0$)  there is a constant $C > 0$ so that
\begin{equation*}
\espalt{}{YV'(yY)1_{yY\geq \hat{y}}} \leq \espalt{}{YV'(Y)1_{yY\geq \hat{y}}}
\leq  C\espalt{}{V(Y)} < \infty.
\end{equation*}
Thus, $\lim_{y\downarrow 0} g(y) = -\infty$ and the result holds.

\end{proof}


\begin{lemma}\label{L:Y_V_ubounds}

Let $\alpha > 0$, $p>1$ and $l>0$. Let $u>0$ and $Y\geq 0$ be such that $\espalt{}{Y} =1$.  Then, for each $0 < \eps < u$ there exists a constant $\ol{C}(\eps,U)>0$
independent of $Y$ and $u$ such that
\begin{equation}\label{E:Y_V_ubounds}
\inf_{y>0}\frac{1}{y}\left(\espalt{}{V(yY)} + u\right) \leq \ol{C}(\eps,U)
+ \begin{cases} \frac{1+\eps}{\alpha}\espalt{}{Y\log(Y)} + u & U\in\Ua\\ \frac{1}{\alpha}\espalt{}{Y\log(Y)} + u & U\in\UUa \\ \left(l(u+\eps)\right)^{1/p}\left((1+\eps) \espalt{}{Y^\gamma}\right)^{1/\gamma} & U\in\Up\end{cases}.
\end{equation}

\end{lemma}

\begin{proof}
Clearly, $\inf_{y > 0} (1/y)\left(\espalt{}{V(yY)} + u\right) \leq \espalt{}{V(Y)} +
u$. Let $\eps > 0$. In view of \eqref{eq: dual_funct_ubounds}, there is some
constant $\ol{C} = \ol{C}(\eps,U)$ so that $\espalt{}{V(Y)} + u\leq \ol{C} +
((1+\eps)/\alpha)\espalt{}{Y\log(Y)} + u$. This proves \eqref{E:Y_V_ubounds}
for $U\in\Ua$. Now, consider $U\in\UUa$ and define
\begin{equation}\label{E:fu_def}
f_U(z) \dfn V(z)-\frac{1}{\alpha}z(\log(z) - 1).
\end{equation}
Using the definition of $\UUa$, calculation shows that
$\limsup_{z\uparrow\infty}\left|f_U(z)\right|/z < \infty$. Since $f_U(0) = 0$, there
is some $M = M(\eps,U)$ so that $f_U(z)\leq M(1+z)$ for $z > 0$. Therefore,
since $\espalt{}{Y}=1$:
\begin{equation*}
\espalt{}{V(Y)} + u = \frac{1}{\alpha}\espalt{}{Y\log(Y)} + \espalt{}{f_U(Y)}
+ u\leq \frac{1}{\alpha}\espalt{}{Y\log(Y)} + 2M + u.
\end{equation*}
Thus, \eqref{E:Y_V_ubounds} holds for $U\in\UUa$. Lastly, for
$U\in\Up$ since $\lim_{z\downarrow 0}V(z) = 0$ and \eqref{E:dual_util_limit_power} holds, for any $0 < \eps < u$ there
is some $M = M(\eps,U)$ so that $V(z) < \eps$ on $z<1/M$, $V(z) \leq M$ on
$1/M \leq z\leq M$ and $V(z) \leq (1+\eps)\hat{l}z^\gamma$ on $z>M$.   Thus, by
splitting $yY$ according to whether $yY < 1/M$, $1/M \leq yY \leq M$ and $yY >
M$ one obtains
\begin{equation*}
\frac{1}{y}\left(\espalt{}{V(yY)} + u\right) \leq \frac{u+\eps}{y} +
V(M)\frac{1}{y}\espalt{}{1_{yY\geq 1/M}} +
(1+\eps)\hat{l}y^{\gamma-1}\espalt{}{Y^\gamma}.
\end{equation*}
Now, $(1/y)\espalt{}{1_{yY\geq 1/M}} \leq M$ since $\espalt{}{Y} =
1$. Additionally, a direct calculation shows that
\begin{equation*}
\inf_{y>0}\left(\frac{u+\eps}{y} +
  (1+\eps)\hat{l}y^{\gamma-1}\espalt{}{Y^\gamma}\right) =
\gamma\left(\frac{u+\eps}{\gamma-1}\right)^{1/p}\left((1+\eps)\hat{l}\espalt{}{Y^\gamma}\right)^{1/\gamma},
\end{equation*}
Since $\gamma(\gamma-1)^{-1/p}\hat{l}^{1/\gamma} = l^{1/p}$,
\eqref{E:Y_V_ubounds} holds for $U\in\Up$.
\end{proof}


\begin{lemma}\label{L:Y_V_lbounds}
Let $\alpha > 0$, $p>1$ and $l>0$. Let $u > 0$ and $Y\geq 0$ be such that $\espalt{}{Y} =
1$. Then for each $0 < \eps < \min\{u,1\}$ there exist constants
  $\ul{C}(\eps,U),\ul{D}(\eps,U)>0$ independent of $Y$ and $u$ such that
\begin{equation}\label{E:Y_V_lbounds}
\inf_{y>0}\frac{1}{y}\left(\espalt{}{V(yY)} + u\right) \geq -\ul{C}(\eps,U)
+ \begin{cases}\frac{1-\eps}{\alpha}\espalt{}{Y\log(Y)} + \ul{D}(\eps,U)\log(u)  & U\in\Ua\\ \frac{1}{\alpha}\espalt{}{Y\log(Y)} + \ul{D}(\eps,U)\log(u) & U\in\UUa \\ \left(l(u-\frac{\eps}{2})\right)^{1/p}\left((1-\eps)\espalt{}{Y^\gamma}\right)^{1/\gamma} & U\in\Up\end{cases}.
\end{equation}

\end{lemma}

\begin{proof}[Proof of Lemma \ref{L:Y_V_lbounds}]
Let $0 < \eps < \min\{u,1\}$. In view of $V(0)= 0$ and \eqref{E:dual_util_limit}, there is
some $M=M(\eps,U)$ large enough so that $V(z)\geq -\eps/2$ on $z<1/M$,
$V(z)\geq U(0)$ on $1/M\leq z \leq M$ and $V(z)\geq
(1-\eps)(1/\alpha)z(\log(z)-1)$ on $z>M$. Since $U(0)<0$, by splitting $yY$
according to whether $yY < 1/M$, $1/M\leq yY\leq M$ or $yY>M$ one obtains
\begin{equation}\label{E:new_Y_V_lbound}
\begin{split}
\frac{1}{y}\left(\espalt{}{V(yY)} +u\right) &\geq \frac{u-\eps/2}{y} + U(0)\frac{1}{y}\espalt{}{1_{1/M\leq yY}} +
\frac{1-\eps}{\alpha}\espalt{}{Y(\log(yY)-1)(1-1_{yY\leq M})}.
\end{split}
\end{equation}
As before $(1/y)\espalt{}{1_{yY\geq 1/M}} \leq
M$. Furthermore, $\espalt{}{Y(\log(yY)-1)1_{yY\leq M}} \leq \log(M)-1$. Thus
\begin{equation*}
\frac{1}{y}\left(\espalt{}{V(yY)}  + u\right)\geq \frac{u-\eps/2}{y} + U(0)M +
\frac{1-\eps}{\alpha}\log(y) + \frac{1-\eps}{\alpha}\espalt{}{Y\log(Y)}  - \frac{1-\eps}{\alpha}\log(M).
\end{equation*}
Since $u-\eps/2 > 0$,
\begin{equation}\label{E:y_var_calc}
\inf_{y>0}\left(\frac{1-\eps}{\alpha}\log(y) + \frac{u-\eps/2}{y}\right) = \frac{1-\eps}{\alpha}\left(1+\log\left(\frac{\alpha(u-\eps/2)}{1-\eps}\right)\right),
\end{equation}
By adding and subtracting $(1-\eps)/\alpha \log(u)$ and using that
$1-\eps/(2u) \geq 1/2$, \eqref{E:Y_V_lbounds} holds for $U\in\Ua$ with
\begin{equation*}
\begin{split}
\ul{C}(\eps,U) &=  \left(U(0)M  +
\frac{1-\eps}{\alpha}\left(1 +
  \log\left(\frac{\alpha}{2M(1-\eps)}\right)\right)\right);\qquad \ul{D}(\eps,U) = \frac{1-\eps}{\alpha}.\\
\end{split}
\end{equation*}
Regarding \eqref{E:Y_V_lbounds} for $U\in\UUa$, let $f_U$ be
as in \eqref{E:fu_def} and recall from the previous lemma that $f(0) = 0$
and $\limsup_{z\uparrow\infty}|f_U(z)|/z < \infty$.  Given this, for the given
$\eps$ there exists constants $M = M(\eps,U)$ and $K = K(\eps,U)$ so that $f_U(z)\geq -\eps/2$ for $z < 1/M$, $f_U(z) \geq -K$ for $1/M\leq z\leq M$
and $f_U(z)\geq -Kz$ on $z>M$. Thus
\begin{equation}\label{E:new_Y_V_lbound_Uac}
\begin{split}
\frac{1}{y}\espalt{}{f_U(yY)} &\geq -\frac{\eps}{2y}
-K\frac{1}{y}\espalt{}{1_{1/M \leq yY}} - K\espalt{}{Y1_{yY>M}}\geq -\frac{\eps}{2y} - K(1+M),
\end{split}
\end{equation}
where the last inequality follows because $(1/y)\espalt{}{1_{1/M\leq yY}} \leq
M$ and $\espalt{}{Y}= 1$. This gives
\begin{equation*}
\begin{split}
\frac{1}{y}\left(\espalt{}{V(yY)} + u\right) & =
\frac{1}{\alpha}\espalt{}{Y(\log(yY)-1)} + \frac{1}{y}\espalt{}{f_U(yY)} + \frac{u}{y},\\
&\geq \frac{u-\eps/2}{y} + \frac{1}{\alpha}\log(y) +
\frac{1}{\alpha}\espalt{}{Y\log(Y)} -\frac{1}{\alpha} - K(1+M).
\end{split}
\end{equation*}
\eqref{E:Y_V_lbounds} follows by repeating the argument begun in \eqref{E:y_var_calc}
above, except that $1/\alpha \log(y)$ replaces
$((1-\eps)/\alpha)\log(y)$. Lastly,
\eqref{E:Y_V_lbounds} for $U\in\Up$is treated. Here, the calculations are the same as
above except that $V(z) \geq (1-\eps)\hat{l} z^\gamma$ on $z > M$.  This gives
\begin{equation*}
\begin{split}
\frac{1}{y}\left(\espalt{}{V(yY)} +u\right) &\geq \frac{u-\eps/2}{y} + U(0)M +
(1-\eps)\hat{l}y^{\gamma-1}\espalt{}{Y^\gamma(1-1_{yY\leq M})}.
\end{split}
\end{equation*}
$y^{\gamma-1}\espalt{}{Y^\gamma 1_{yY\leq M}} \leq M^{\gamma-1}$ since
$\espalt{}{Y}=1$. This gives
\begin{equation*}
\begin{split}
\frac{1}{y}\left(\espalt{}{V(yY)} +u\right) &\geq \frac{u-\eps/2}{y} + U(0)M - (1-\eps)\hat{l}M^{\gamma-1}+
(1-\eps)\hat{l}y^{\gamma-1}\espalt{}{Y^\gamma}.
\end{split}
\end{equation*}
Calculation shows that
\begin{equation*}
\inf_{y>0}\left(\frac{u-\eps/2}{y} + (1-\eps)\hat{l}
  y^{\gamma-1}\espalt{}{Y^\gamma}\right) =
\gamma\left(\frac{u-\eps/2}{\gamma-1}\right)^{1/p}\left((1-\eps)\hat{l}\espalt{}{Y^\gamma}\right)^{1/\gamma}.
\end{equation*}
Since as shown in the previous lemma,
$\gamma(\gamma-1)^{-1/p}\hat{l}^{1/\gamma} = l^{1/p}$,
\eqref{E:Y_V_lbounds} holds with $\ul{C}(\eps,U) = -U(0)M +
(1-\eps)\hat{l}M^{\gamma-1}$.
\end{proof}

%


\bibliographystyle{siam}
\bibliography{master}

\end{document}